\documentclass[fleqn,usenatbib]{mnras}
\usepackage{newtxtext,newtxmath}
\usepackage[T1]{fontenc}

\DeclareRobustCommand{\VAN}[3]{#2}
\let\VANthebibliography\thebibliography
\def\thebibliography{\DeclareRobustCommand{\VAN}[3]{##3}\VANthebibliography}

\newcommand{\overbar}[1]{\mkern 1.5mu\overline{\mkern-1.5mu#1\mkern-1.5mu}\mkern 1.5mu}
\def\siglos{\sigma_{\text{LOS}}}

\def\dd{\text{d}}

\usepackage{graphicx}
\usepackage{amsmath}
\usepackage{amssymb}
\usepackage{dsfont}
\usepackage[many]{tcolorbox}
\usepackage{multirow}

\title[Fermionic DM and dSph Galaxies]{New Constraints on the Mass of Fermionic Dark Matter from Dwarf Spheroidal Galaxies}

\author[Alvey et al.]{James Alvey,$^{1}$\thanks{james.alvey@kcl.ac.uk}
Nashwan Sabti,$^{1}$\thanks{nashwan.sabti@kcl.ac.uk}
Victoria Tiki,$^{2}$\thanks{tiki@lorentz.leidenuniv.nl}
Diego Blas,$^{1}$
Kyrylo Bondarenko,$^{3, 4}$
\newauthor
Alexey Boyarsky,$^{2}$
Miguel Escudero,$^{1,5}$
Malcolm Fairbairn,$^{1}$
Matthew Orkney$^{6}$
\newauthor
and Justin I.  Read$^{6}$
\vspace{15pt}
\\
$^{1}$Department of Physics, King’s College London, Strand, London WC2R 2LS, UK\\
$^{2}$Intituut-Lorentz, Leiden University, Niels Bohrweg 2, 2333 CA Leiden, The Netherlands\\
$^{3}$Theoretical Physics Department, CERN, 1 Esplanade des Particules, Geneva 23, CH-1211, Switzerland\\
$^{4}$L’Ecole Polytechnique F\'{e}d\'{e}rale de Lausanne, Route Cantonale, 1015 Lausanne, Switzerland\\
$^{5}$Physik-Department, Technische Universit{\"{a}}t, M{\"{u}}nchen, James-Franck-Stra{\ss}e, 85748 Garching, Germany\\
$^{6}$Department of Physics, University of Surrey, Guildford, GU2 7XH, UK
}

\date{\vspace{0pt}}

\pubyear{2020}

\begin{document}
\label{firstpage}
\pagerange{\pageref{firstpage}--\pageref{lastpage}}
\maketitle

\begin{abstract}
Dwarf spheroidal galaxies are excellent systems to probe the nature of fermionic dark matter due to their high observed dark matter phase-space density. In this work, we review, revise and improve upon previous phase-space considerations to obtain lower bounds on the mass of fermionic dark matter particles. The refinement in the results compared to previous works is realised particularly due to a significantly improved Jeans analysis of the galaxies. We discuss two methods to obtain phase-space bounds on the dark matter mass, one model-independent bound based on Pauli's principle, and the other derived from an application of Liouville's theorem. As benchmark examples for the latter case, we derive constraints for thermally decoupled particles and (non-)resonantly produced sterile neutrinos. Using the Pauli principle, we report a model-independent lower bound of $m \geq 0.18\,\mathrm{keV}$ at 68\% CL and $m \geq 0.13\,\mathrm{keV}$ at 95\% CL. For relativistically decoupled thermal relics, this bound is strengthened to $m \geq
0.59\,\mathrm{keV}$ at 68\% CL and $m \geq 0.41\,\mathrm{keV}$ at 95\% CL, whilst for non-resonantly produced sterile neutrinos the constraint is $m \geq 2.80\,\mathrm{keV}$ at 68\% CL and $m \geq 1.74\,\mathrm{keV}$ at 95\% CL. Finally, the phase-space bounds on resonantly produced sterile neutrinos are compared with complementary limits from X-ray, Lyman-$\alpha$ and Big Bang Nucleosynthesis observations.
\end{abstract}

\begin{keywords}
cosmology: dark matter --- galaxies: dwarf\\

\vspace*{-5pt} \noindent \href{https://github.com/james-alvey-42/FermionDSph}{\raisebox{-1pt}{\includegraphics[width=9pt]{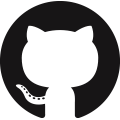}}}\hspace{2pt} Tabulated results, code for computing the bounds and dwarf profiles can be found \href{https://github.com/james-alvey-42/FermionDSph}{here}.
\end{keywords}

\section{Introduction}
\label{sec:introduction}
The astrophysical and cosmological evidence for dark matter is overwhelming. Precise observations of the Cosmic Microwave Background~\citep{Aghanim:2018eyx} show that 26\% of the energy density of the Universe consists of a form of matter that appears to interact only gravitationally. In addition, innumerable studies suggest strongly that galaxies, clusters of galaxies, and the vast majority of the virialised objects in the Universe are dominated by some matter that does not emit light -- see e.g.~\cite{Rubin:1970zza,Peebles:1984zz,Dodelson:2003ft,Clowe:2006eq, Dodelson:2011qv, Pardo:2020epc}.

The nature of dark matter represents one of the major mysteries of modern physics. As such, a vast number of dark matter candidates have been put forward. For instance, in terms of their mass, the landscape of dark matter candidates extends from $m \sim 10^{-22}\,\text{eV}$ for ultralight dark matter~\citep{Hui:2016ltb} to $m \sim 100 \, M_\odot$ in the case of primordial black holes~\citep{Carr:2016drx}. In between, there is an array of well-motivated dark matter candidates, such as WIMPs with $m \sim 1\,\text{MeV}-100\,\text{TeV}$~\citep{Bertone:2004pz}, axions~\citep{Marsh:2015xka}, sterile neutrinos with $ m \sim(1-100)\,\text{keV}$~\citep{Boyarsky:2018tvu}, and asymmetric dark matter with $ m \sim 1\, \text{GeV}$~\citep{Zurek:2013wia}. The majority of well-motivated dark matter candidates were proposed decades ago, and as a result, an exhaustive experimental programme has been developed to search for them. At this point in time, however, all direct searches for dark matter have returned a null result. 

Laboratory experiments are not the only way to probe the nature of dark matter. In fact, astrophysics and cosmology have long been used to constrain its properties. A pioneering astrophysical test of fermionic dark matter was carried out in~\cite{Tremaine:1979we}. They used the Pauli exclusion principle together with the fact that dwarf galaxies have a high abundance of dark matter to place a lower bound on its mass. At the time, the Tremaine and Gunn bound was used to disfavour the possibility that active neutrinos were the dark matter of the Universe. We now know that neutrinos cannot constitute both the astrophysical and cosmological dark matter, but nonetheless the argument still applies for any generic fermionic dark matter particle. 

Indeed, there is still plenty of motivation for relatively light fermionic dark matter candidates, see e.g.~\cite{Boyarsky:2018tvu} for a review. This motivation primarily arises from the fact that keV-scale sterile neutrinos can be produced in the early Universe in the right abundance to constitute the entirety of the dark matter, e.g.~\cite{Dodelson:1993je, Shi:1998km}. In addition, such particles can be embedded into frameworks explaining the origin of neutrino masses and the observed matter-antimatter asymmetry of the Universe~\citep{Asaka:2005pn}. 

Given this, a variety of astrophysical and cosmological tests have been considered to constrain the masses of sterile neutrinos and other similar dark matter particles. These species fall often within the category of so-called warm dark matter, because they typically possess some residual kinetic energy at the epoch of galaxy formation. There are three main sets of bounds that keV-scale fermionic dark matter particles should satisfy:

\begin{itemize}
    \item \textit{Phase-space constraints ---} As envisaged by~\cite{Tremaine:1979we}, there is a limit to the extent to which fermionic dark matter particles can be compressed within a halo -- see e.g.~\cite{Dalcanton:2000hn,Boyarsky:2008ju}.  This class of constraints applies to all keV-scale fermionic dark matter particles whether they are warm dark matter or not.
    
    \item \textit{Galaxy count constraints ---} Warm dark matter suppresses the growth of structure below its free-streaming length (on the scale of smaller galaxies). One should then ensure that it is possible to at least form the Milky Way satellites we observe -- see e.g.~\cite{Anderhalden:2012jc, Lovell:2015psz, Read:2017lvq,Kim:2017iwr,Jethwa:2016gra,Nadler:2020prv}.
    
    \item \textit{Lyman-$\alpha$ forest constraints ---} Warm dark matter suppresses the growth of structure at small scales, yet all observations of the Lyman-$\alpha$ forest appear to be compatible with the expectations of cold dark matter -- see e.g.~\citep{Efstathiou:2000kk,Viel:2013fqw,Garzilli:2019qki,Palanque-Delabrouille:2019iyz}. 
\end{itemize}

Note that in the case where the dark matter particles can decay into photons (e.g. sterile neutrinos), X-ray constraints could further constrain their parameter space, see e.g.~\cite{Abazajian:2001vt, Boyarsky:2009ix} for a review. Other promising techniques which may further constrain the warm dark matter mass include observations of stellar streams, e.g.~\cite{Ibata:2001iv, Johnston:2001wh, Banik:2019smi, Helmi:2020otr}, and gravitational lensing systems, e.g.~\cite{Bacon:2009aj, Vegetti:2018dly,Gilman:2019nap}.\\

In this work, we apply phase-space arguments to dwarf spheroidal galaxies in order to put a lower bound on the fermionic dark matter mass. There are two main reasons to do this: Firstly, bounds based on phase-space arguments are largely insensitive to the cosmological setting, instead being based on robust statistical theorems or principles. Secondly, on the data side, we benefit from an improved kinematical analysis of dwarf spheroidal galaxies from~\cite{Read:2018fxs}, which allows us to better determine the density profile. We study two different ways of obtaining such a bound, each strengthening upon the previous one:

\begin{itemize}
    \item \textit{Pauli Exclusion Principle} --- We first derive a robust, model-independent bound. For this, we consider a self-gravitating object that fully consists of a degenerate Fermi gas and require that the Fermi velocity does not exceed the escape velocity.
    
    \item \textit{Liouville's Theorem} --- A stronger, but model-dependent, bound can be obtained by using Liouville's theorem, which implies that in a collisionless and dissipationless evolution, the maximum of the dark matter phase-space distribution inside the dwarf galaxy cannot exceed the primordial value.
    In order to apply Liouville's theorem, the form of the primordial distribution needs to be known. This is clearly model-dependent and as such we consider a number of important benchmark cases. These include \emph{(i)} relativistically decoupled thermal dark matter and \emph{(ii)} sterile neutrino dark matter (both non-resonantly and resonantly produced). We also provide a prescription for how such bounds can be obtained for any other fermionic dark matter candidate.
\end{itemize}

Pivotal to this approach is the knowledge of the dark matter phase-space density within the dwarf galaxy itself. We note that several studies have dealt with similar bounds, e.g.~\cite{Dalcanton:2000hn,Boyarsky:2008ju,Domcke:2014kla,DiPaolo:2017geq,Savchenko:2019qnn}, however, we supersede these thanks to an improved Jeans analysis. By making use of higher order moments of the line-of-sight stellar velocity distribution, we are able to better measure the mass (and therefore density) of the dark matter within the galaxy. We improve the Liouville analysis in two stages, which will eventually lead to two bounds, the second being stronger than the first but containing slightly more assumptions. Note that the first stage also applies to the method using the Pauli exclusion principle.

The first stage is simply to calculate the density and escape velocity as a function of radius to place a lower limit on the phase-space density. This is based on the fact that all dark matter located at a given radius must have a velocity less than the escape velocity at that radius. This will give a stronger constraint than the ones in previous studies, see particularly~\cite{Boyarsky:2008ju}, because in this reference the phase-space density was evaluated at the half-light radius. In this work, with the improved Jeans modelling, however, we can go to lower radii, where we will find that the phase-space density is higher.

The second stage is to note that the assumption that the dark matter fills the whole of velocity volume in phase space evenly all the way to the escape velocity \emph{underestimates} the phase-space density. The velocity distribution is typically not uniform and can be larger for velocities less than the velocity dispersion. Indeed this can be a significant effect depending on the value of the dark matter velocity anisotropy $\beta_\mathrm{dm}(r)$. In order to estimate the velocity dispersion of dark matter, we need to solve the Jeans equation \emph{again}, this time for the dark matter particles, with some realistic range of $\beta_\mathrm{dm}(r)$. Following this procedure we can constrain the dark matter velocity dispersion. This is a novel approach and allows us to obtain a more accurate estimate of the phase-space density.

The remainder of this paper is organised as follows: In Sec.~\ref{sec:JeansAnalysis} we provide a quick recap of the standard Jeans analysis and explain our improved approach. Sec.~\ref{sec:HowToObtainBounds} then lays out a detailed prescription on how the phase-space bounds are obtained in this work. We then apply this formalism on a set of benchmark models in Sec.~\ref{sec:benchmark_models}. In Sec.~\ref{sec:Results} we present our results and a detailed comparison with previous studies. Finally, we draw our conclusions in Sec.~\ref{sec:Conclusions}. 

\section{Improved Jeans Analysis}
\label{sec:JeansAnalysis}

Dwarf spheroidal galaxies are faint, low mass objects, usually thought to exist around larger host galaxies. Currently, only those around the Milky Way and Andromeda can be detected using telescopes, due to their extremely low luminosity~\citep{Simon2019xd}. These galaxies have a very large mass-to-light ratio, implying that they are almost completely made out of dark matter. They are thought to form within subhalos that collapse at higher redshift than large galaxies like the Milky Way~\citep{White:1994bn}. The early collapse time of these objects results in a large central density of dark matter $\rho_\mathrm{dm}$, while their small overall virial mass implies a low velocity dispersion $\sigma$. This means that the phase-space density of dark matter in these objects (${\sim}\rho_\mathrm{dm}/\sigma^3$) is naturally one of the highest observable in the local Universe.

We are able to better determine the density of dark matter in the inner regions of the dwarf galaxy, far within the half-light radius, by using a new Jeans analysis based on the work in \cite{Read:2018fxs}. In this part of the dwarf, the phase-space density is at its largest, so accessing it naturally leads to improved limits on the mass of fermionic dark matter. In what follows, we will first summarise the formalism of the Jeans method and then detail the improved approach utilised to obtain the dark matter density.

\subsection{Jeans equation}
\label{subsec:jeans_review}
The Jeans equation~\citep{1922MNRAS..82..122J} is obtained from the collisionless Boltzmann equation by assuming a steady-state solution together with spherical symmetry. The traditional approach is to take the second moments of the 6D distribution function $f(\vec{x},\vec{v})$, giving rise to:
\begin{equation}
   \frac{1}{\nu}\frac{\partial}{\partial r}\left(\nu\sigma_r^2\right)+2\frac{\beta\sigma_r^2}{r}=-\frac{GM(r)}{r^2}\ ,
   \label{eqn:jeans}
\end{equation}
where $\nu(r)$ is the spherically averaged tracer density and $\beta(r)$ is the velocity anisotropy:
\begin{equation}
\beta = 1 - \frac{\sigma_t^2}{\sigma_r^2}\ .
\label{eqn:beta}
\end{equation}
Here, $\sigma_t$ and $\sigma_r$ are the tangential and radial velocity dispersions, respectively. Note that in this paper we will be referring to $\beta$ in the context of the velocity distribution of stars, while we will use $\beta_\mathrm{dm}$ for the same quantity for dark matter.

Solutions to the Jeans equation~\eqref{eqn:jeans} 
are subsequently used to determine the line-of-sight velocity dispersion, given by \citep{1982MNRAS.200..361B}:
\begin{equation}
 \siglos^2(R) = \frac{2}{\Sigma(R)}\int_R^\infty \left(1\!-\!\beta\frac{R^2}{r^2}\right)
    \nu\sigma_r^2\,\frac{r\,\dd r}{\sqrt{r^2\!-\!R^2}} \ ,
    \label{eqn:LOS}
\end{equation}
where $\Sigma(R)$ denotes the tracer surface mass density at projected radius $R$ (2D surface density flattened onto a plane).
In order to reconstruct the density of dark matter in the inner regions of the galaxy, the goal of the exercise is to go from observations of $\siglos(R)$ to the mass $M(r)$ as a function of radius. It is clear, however, when examining Eqs.~\eqref{eqn:jeans} and \eqref{eqn:LOS} that different choices of the stellar velocity anisotropy $\beta(r)$ will lead to different conclusions about $M(r)$ for the same $\siglos(R)$.
Since we are unable to obtain information about the proper motion (and therefore $\beta$) of stars at the distance of dwarf galaxies around the Milky Way, this results in a well-known and real problem, sometimes referred to as the $\beta$-degeneracy problem\footnote{This is of course motivation for proposed telescopes like Theia which might be able to measure proper motions for such distant stars~\citep{Boehm:2017wie}.}.

One discovery that has been made is that this degeneracy is broken at the half-light radius. This means that regardless of the choice of $\beta(r)$, the mass enclosed at that radius does not vary significantly~\citep{Wolf:2009tu}. Several groups have used this to their benefit by identifying disparate groups of stars within the same halo separated by their metallicity. They then used these separate populations to obtain more robust mass estimates at different radii and consequently improve their determination of the dark matter density profile~\citep{Amorisco:2011hb,Walker:2011zu}\footnote{It should be noted that recent numerical experiments have opened some questions about this method~\citep{2018MNRAS.474.1398G}.}. In this work, we will use a different approach to break the $\beta$-degeneracy, which is the topic of the next section.

\subsection{Higher Moments of the velocity dispersion}
\label{subsec:higher_order_moments}
There is an alternative when looking to break this problematic degeneracy, which  involves considering higher moments of the line-of-sight velocity distribution --- specifically the fourth moments. This provides complementary information to just the dispersion, i.e. the second-order moment. In particular, \cite{1990AJ.....99.1548M} constructed virial estimators based on both the second-order velocity dispersions and the fourth-order moments\footnote{Interestingly, it turned out that using only the second-order estimators was almost as constraining as the traditional Jeans analysis.}. They showed that these observables could be written in terms of the quantities $\beta(r)$, $M(r)$, and $\sigma_r(r)$ and therefore could be used as additional data to restrict solutions to the spherical Jeans equation. Explicitly, these fourth-order moments are given by:
\begin{equation}
\int^{\infty}_{0} \Sigma \langle v^4_{LOS} \rangle R \, \rm{d} R = \frac{2}{5} \int^{\infty}_{0} \nu (5-2\beta) \sigma_r^2 GM R \, \rm{d}R \ ,
\label{eqn:vs1}
\end{equation}
\begin{equation}
\int^{\infty}_{0} \Sigma \langle v^4_{LOS} \rangle R^3 \, \rm{d} R = \frac{4}{35} \int^{\infty}_{0} \nu (7-6\beta) \sigma_r^2 GM R^3 \, \rm{d}R \ .
\label{eqn:vs2}
\end{equation}
These estimators have been applied in a number of settings, most notably to the analysis of spherical galaxies~\citep{Gerhard:1993tn}, as well as to the determination of the dark matter distribution at the scale of dwarfs and clusters~\citep{Lokas:2003ks, Lokas:2004sw}. The ability of these higher order moments of the velocity distribution to constrain the internal dynamics of dwarf galaxies was studied in \cite{Richardson:2012ig, Richardson:2014mra}, where the authors found evidence for a cusp in the density profile of the Sculptor galaxy.

\begin{figure}
    \centering
    \includegraphics[width=\linewidth]{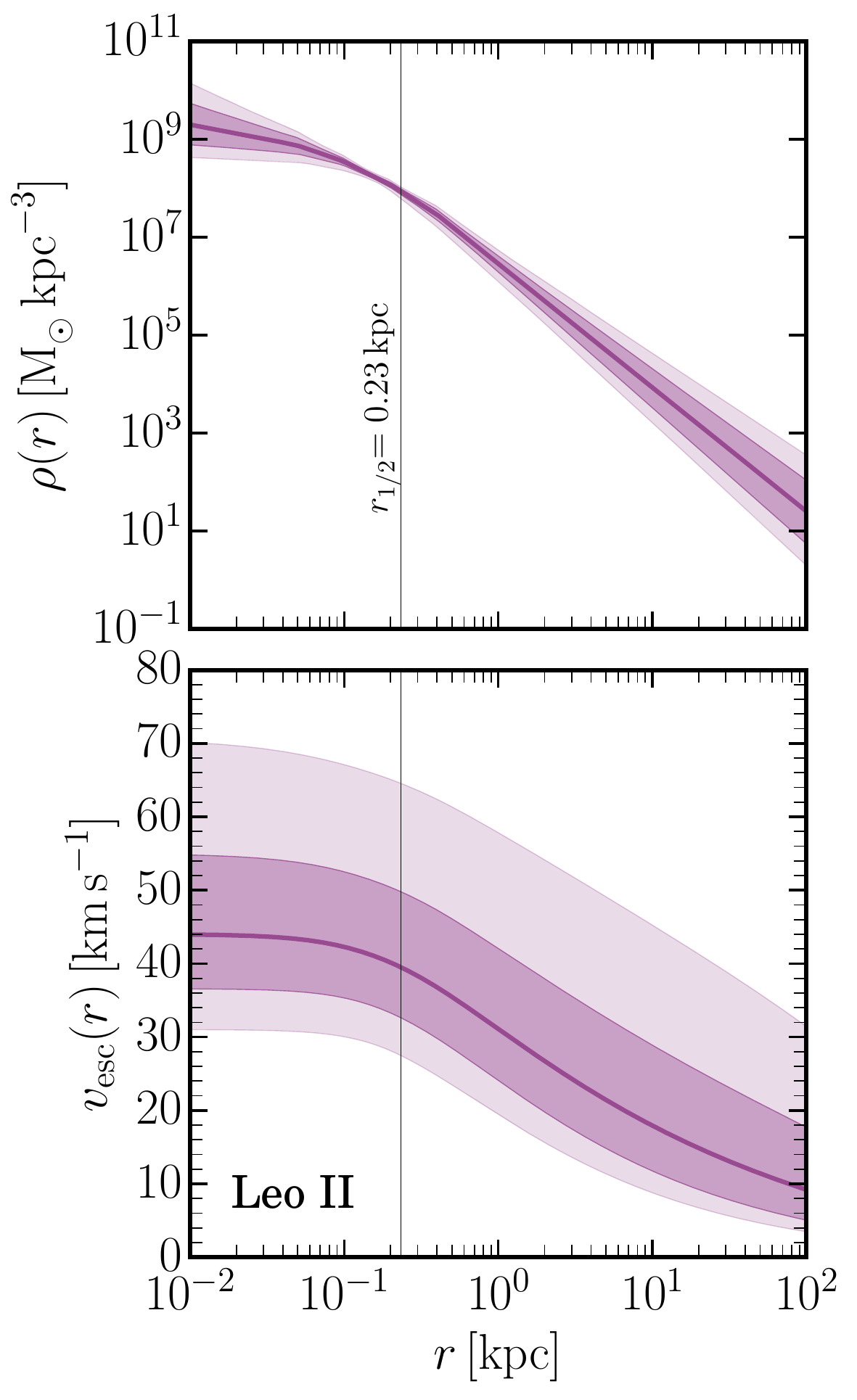} \\
    \caption{\emph{Upper panel:} The density profile of the dwarf galaxy Leo II derived using the method described in Sec.~\ref{subsec:higher_order_moments}. \emph{Lower panel:} The escape velocity as a function of radius of the same galaxy obtained using Eq.~\eqref{eqn:vesc}. In both panels, the solid line refers to the average value of the quantity whilst the darker and lighter shaded regions are the 68\% and 95\% CL regions respectively. The value $r_{1/2}$ refers to the 3D projected half-light radius of the galaxy, where the uncertainty is expected to be minimal.}
    \label{fig:properties_Leo2}
\end{figure}

Most relevant to this current work is the fact that Eqs.~\eqref{eqn:vs1} and~\eqref{eqn:vs2} were subsequently included in a new code, \texttt{Gravsphere}~\citep{2018MNRAS.481..860R}. There are a number of interesting features to this work, including the possibility of constraining a non-parametric density function. As noted above, breaking the $\beta$-degeneracy means that the density within the inner part of the dwarf can be far better constrained. The \texttt{Gravsphere} code was applied to new data for the classical dwarf galaxies in \cite{Read:2018fxs}, where the authors came to the interesting conclusion that there seems to be a correlation between star formation and cored dark matter profiles. In our work, we make use of the same data as in this reference, which includes photometric data from~\cite{Koposov:2014lja, McMonigal:2014naa, flewelling2019panstarrs1} and stellar-kinematic data from~\cite{Mateo:2007xh, Walker:2008ax, Walker_2015, Spencer_2017, Spencer_2018}, in addition to data from~\cite{Simon:2007dq,McConnachie:2012vd, Alvarez:2020cmw}. We expect that the galaxies with the most cusp-like profiles will have the largest dark matter phase-space densities and are therefore the objects of most significance to the bound presented in this work.

The results of using the \texttt{Gravsphere} code for the dwarf galaxy Leo II are shown in Fig.~\ref{fig:properties_Leo2}, along with the derived escape velocity as a function of the radius within the galaxy. By itself, this is enough information to derive a robust constraint on the dark matter mass using the methodology described in Secs.~\ref{subsec:bounds_from_pauli_method} and~\ref{sec:maxcoarsegraining}. Nonetheless, as we noted above, due to the small virial mass we expect that the velocity dispersion within these systems to be small compared to the escape velocity. Therefore, as described in Sec.~\ref{sec:gausscoarsegraining}, if we can combine the new information about the density profile with some determination of the velocity dispersion of the dark matter, then we can expect the bounds on the mass of the dark matter to further improve. For this purpose, we solve a second spherical Jeans equation, but this time for the dark matter:
\begin{equation}
   \frac{1}{\rho}\frac{\partial}{\partial r}\left(\rho \,\sigma_{r, \mathrm{dm}}^2\right)+2\frac{\beta_{\rm dm}\sigma_{r, \mathrm{dm}}^2}{r}=-\frac{GM(r)}{r^2}\ ,
   \label{eqn:dmjeans}
\end{equation}
where $\rho \equiv \rho_\mathrm{dm}$ is the density profile for the dark matter, $\sigma_{r, \mathrm{dm}}$ is the dark matter velocity dispersion, and $\beta_{\rm dm}$ is the velocity anisotropy of the dark matter particles. Now, having solved the Jeans equation~\eqref{eqn:jeans} for the stars, we know the mass profile $M(r)$. This can then be plugged in Eq.~\eqref{eqn:dmjeans}, together with an appropriate choice of the dark matter velocity anisotropy, to obtain a solution for the dark matter velocity dispersion as a function of radius. There is a final important point to note when carrying out this analysis: we cannot measure $\beta_{\mathrm{dm}}$, so in order to solve Eq.~\eqref{eqn:dmjeans}, we must either motivate it theoretically, or obtain a prior on $\beta_{\rm dm}(r)$ from N-body simulations of this type of galaxy. In the next section, we justify our choice using a combination of these ideas.

\subsection{Priors on the $\beta_\mathrm{dm}$ profile}\label{subsec:DM_Jeans}

The choice as to which $\beta_{\rm dm}(r)$ should be used to solve for the dark matter velocity dispersion in Eq.~\eqref{eqn:dmjeans} ultimately needs to be guided by theoretical considerations and N-body simulations of dark matter halos. In particular, one should not use the same $\beta$ profile as in Eq.~\eqref{eqn:jeans}, since there is little reason to assume that the dark matter follows exactly the same dynamics as the stars. 

\begin{figure}
    \centering
    \includegraphics[width=\linewidth]{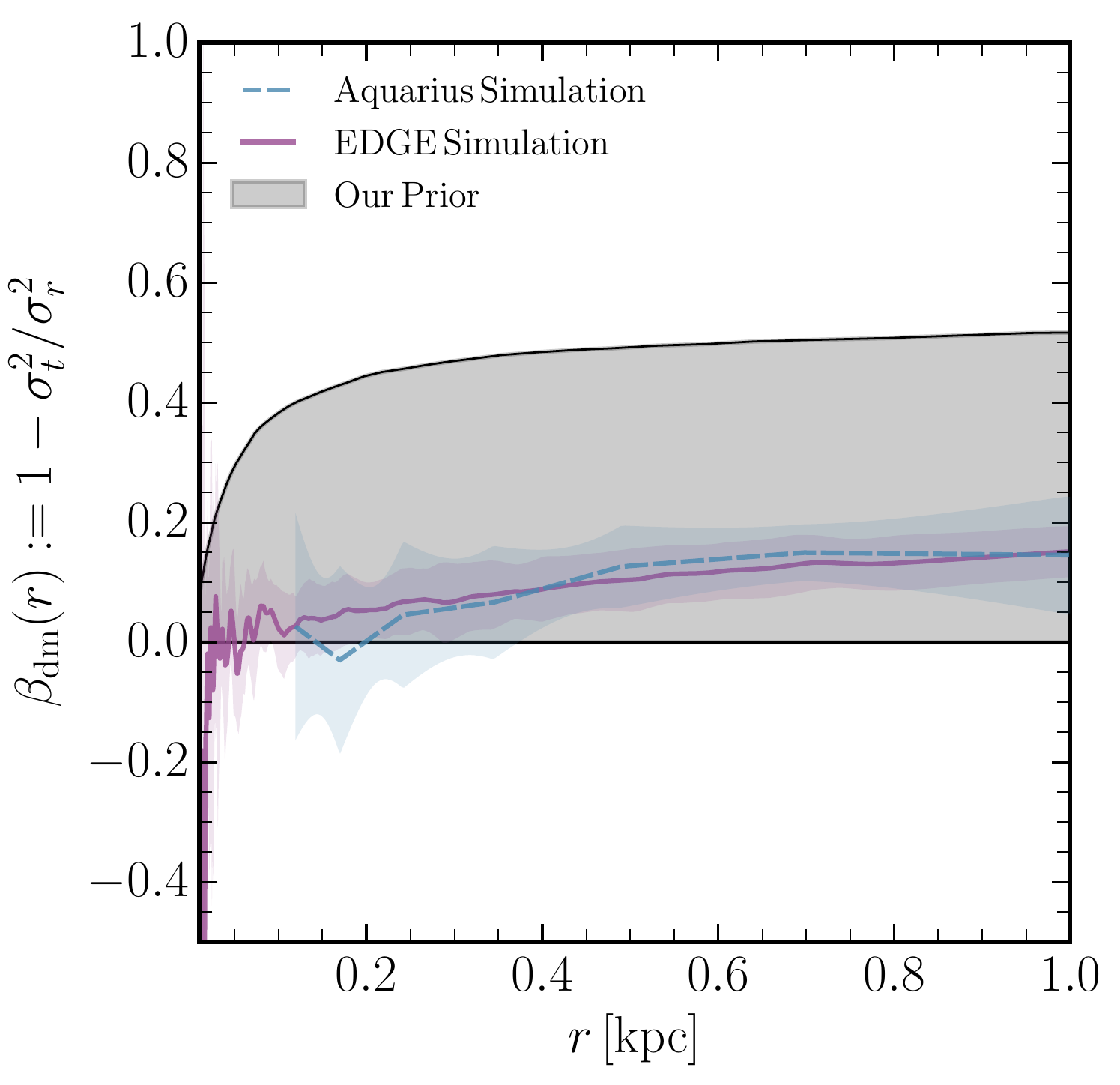}
    \vspace{-6pt}
    \caption{The prior on the $\beta_{\rm dm}$ profile used to solve Eq.~\eqref{eqn:dmjeans} (shown in grey) along with ensemble average profiles derived from the \texttt{EDGE} (shown in purple) and \texttt{Aquarius} (shown in blue) simulations with CDM.\vspace{-5pt}}
    \label{fig:prior_beta}
\end{figure}

At radii larger than ${\sim}0.1\,\mathrm{kpc}$, the prior we choose is based upon $\beta$ profiles derived from the dark matter in the \texttt{EDGE} simulation suite~\citep{2020MNRAS.491.1656A}. \texttt{EDGE} uses the dual N-body and hydrodynamics code \texttt{RAMSES}~\citep{2002A&A...385..337T} to run cosmological zoom simulations of dwarf galaxies. The suite covers a range in halo mass at $z=0$ of $1.5\times10^9 < M/M_{\odot} < 8\times10^9$ at high resolution\footnote{Approximately $120 M_{\odot}$ for dark matter particles and 3 pc for the spatial grid at the highest refinement level.}. We base our $\beta_\mathrm{dm}(r)$ profiles on the dark matter component of the target halo, out to the virial radius $r_{200}$ (the radius at which the galactic density falls to 200 times the critical density of the Universe). The prior on $\beta_\mathrm{dm}$ used in solving the Jeans equation in Eq.~\eqref{eqn:dmjeans}, together with those from the \texttt{EDGE}~\citep{2020MNRAS.491.1656A} and \texttt{Aquarius}~\citep{Springel:2008cc} simulation suites, is shown in Fig.~\ref{fig:prior_beta}.

One might suspect that the velocity anisotropy would be modified when the dark matter becomes warm, which is something expected in many but not all situations where dark matter is made out of light fermions. To test that this is not the case, we also looked at beta profiles from warm dark matter simulations of similar mass halos in the \texttt{Aquarius} suite\footnote{We thank Mark Lovell and Andrew Robertson for providing us with $\beta_\mathrm{dm}$ profiles from \texttt{Aquarius} WDM simulations.}~\citep{Lovell:2013ola}. We found no significant difference in the beta profile as a function of radius (sometimes $\beta_\mathrm{dm}$ becomes very slightly negative). Later we will find that relaxing the priors on the velocity anisotropy parameter much more than what is required to envelope the results from WDM simulations has a relatively small effect on the results.

Near the galactic centre, the simulations quickly deteriorate due to insufficient particle statistics. Nevertheless, there is a theoretical motivation as to why the prior should approach zero towards the centre, i.e., why the inner region is expected to be isotropic. From a theoretical point of view, tangential anisotropy is unusual. Dark matter halos form from roughly cold collapse initial conditions in any reasonable cosmology. This means that they start out as an extended, low density, nearly homogeneous Lagrangian region prior to collapse, with initial relative velocities much smaller than their final relative velocities after virialisation. In this situation, particles collapse nearly radially, generating angular momentum only through weak tidal torques or through the radial orbit instability. The process of virialisation leads to isotropy in the centre, but further out relaxation is incomplete and the halo becomes frozen in a pseudo-equilibrium state. This all leads us to expect approximate isotropy in the centre and radial anisotropy further out.

As mentioned before, we have performed a run with a wider prior on $\beta_{\mathrm{dm}}$ (a factor of $5-10$ times wider near the centre of the dwarf and ${\sim}1.5$ times wider at larger radii). We found, as expected, that the inferred phase-space density mostly changed at small radii, with the central value changing little. This indicates that the modelling is consistent across the different possible choices for $\beta_{\mathrm{dm}}$. As a result, we found no significant change in the central values of our constraints, while  the $1\sigma$ and $2\sigma$ errors increased by at most a factor of $1.2$ and $1.6$ respectively. This shows that our choice for the prior in Fig.~\ref{fig:prior_beta} results in robust constraints, with well-controlled errors.

\vspace{-8pt}\section{How to obtain phase-space bounds on the dark matter mass}
\label{sec:HowToObtainBounds}
In this section we will explain the methodology, assumptions and theoretical considerations that allow us to place a bound on the dark matter mass. The first bound we consider is based on the Pauli exclusion principle, and is independent of both the baryonic physics within the galaxy and the primordial production mechanism, considering only the existence of a self-gravitating halo.

To go further, we must consider more carefully the principles governing the evolution of the primordial distribution function. In particular, we use the fact that for a collisionless, dissipationless system, Liouville's theorem tells us that the maximum value of the phase-space distribution cannot increase. In practice, this means that we must consider a specific production scenario, such as that of a thermal particle, or (non-)resonant sterile neutrino generation. 

This is not the end of the story, however --- in order to set a bound using Liouville's theorem, we need to somehow probe the final state of the distribution function within the dwarf galaxy. This is where the improved Jeans analysis described above plays a crucial role. There are various degrees to which we can try to utilise this analysis to improve our estimate of the final observed phase-space density. We present two methods to make such an estimate, one based on maximal coarse-graining and the other using information about the velocity dispersion of dark matter inside the dwarfs.

\subsection{Bounds from Pauli's principle}
\label{subsec:bounds_from_pauli_method}
This bound is based on Pauli's exclusion principle and considers a self-gravitating object that fully consists of a degenerate Fermi gas\footnote{Note that this work considers the dark matter to constitute a single particle species. The assumption of multi-component dark matter may relax the bounds obtained here~\citep{Davoudiasl:2020uig}.}. The fermions fill up the available states and have a characteristic (Fermi) velocity that is solely determined by the mass and number density of the particles. Such a bound object can only exist if the Fermi velocity does not exceed the escape velocity. A constraint obtained in this way is independent of any primordial production mechanism and does not require any information on the subsequent evolution of the particles. Assuming a fermionic particle with $g$ internal degrees of freedom inside a halo with mass density $\rho(r)$, the Fermi velocity $v_{\mathrm{F}}$ is given by:
\begin{align}
    v_\mathrm{F}(r) = \left(\frac{6\pi^2\rho(r)}{gm^4}\right)^{1/3}\ .
\end{align}
Next, by considering how much work needs to be done to move a particle from a radius $r$ inside the halo to infinity, we find the escape velocity to be:
\begin{align}\label{eqn:vesc}
    v_\mathrm{esc}(r) = \left(8\pi G\int_r^{r_\mathrm{max}}\frac{\mathrm{d}x}{x^2}\int_0^x \rho(y)y^2\mathrm{d}y +  \frac{2GM(<r_\mathrm{max})}{r_\mathrm{max}}\right)^{1/2}\ .
\end{align}

In deriving this expression, we have assumed a cut-off scale $r_\mathrm{max}$ in the density profile, with $M(<\hspace{-0.8mm}r_\mathrm{max})$ denoting the total mass contained within this radius. In this work $r_\mathrm{max}$ represents the outer radius at which the Jeans modelling is performed and is typically around 500 kpc, well beyond the virial radius of the dwarf galaxies considered.\footnote{We choose the radius of 500 kpc to be as robust as possible, and fully consistent with the Jeans modelling. Even if we had used the tidal radius of $\mathcal{O}(1)\,\mathrm{kpc}$, the bounds would have improved by only a factor of ${\sim}1.5$.} The bound can then be obtained by imposing the following condition:
\begin{equation}
    \label{eq:pauli_bound_condition}
    \boxed{v_\mathrm{F}(r) \leq v_\mathrm{esc}(r) \longrightarrow m_\mathrm{deg} \geq \left(\frac{6\pi^2\rho(r)}{gv_\mathrm{esc}(r)^{3}}\right)^{1/4}\ .}
\end{equation}
\noindent This sets a robust lower bound on the dark matter mass. It is worth noting that the derivation of the escape velocity assumes that the halo is spherical. This is investigated quantitatively in Appendix A of \cite{Boyarsky:2008ju}, and is shown to change the results by at most ${\sim}10\%$ for these systems.
\subsection{Bounds from Liouville's theorem}\label{sec:LiouvilleTheorem}
Constraints obtained via the previous method can be further strengthened by using Liouville's theorem. For a collisionless and dissipationsless particle species, this theorem states that the time evolution of the fine-grained distribution function of a particle follows a Hamiltonian flow. In particular, this means that its maximum cannot increase throughout the cosmological evolution~\citep{2008gady.book.....B}. Hence, Liouville's theorem is a useful tool to connect the primordial phase-space density to the one observed today.

In principle, there are two parts to consider: the primordial phase-space density and the one observed in the dwarf. There is a complication in that we do not observe the fine-grained phase-space density at late times. We can only estimate a coarse-grained phase-space density $F^{\mathrm{late}}_{\mathrm{coarse}}$, which is an \emph{average} quantity and thus nonetheless satisfies $F^{\mathrm{late}}_{\mathrm{coarse}} \leq F^{\mathrm{late}}_{\mathrm{fine}}$. Acknowledging this, we can derive a bound on the dark matter mass by writing Liouville's theorem in the following form\footnote{Here, we denote the phase-space density by $F$ and the distribution function by $f$. For a non-relativistic particle, they are related via $F = m^4 f$, since $f = \mathrm{d}n/\mathrm{d}^3p = \mathrm{d}(\rho/m)/\mathrm{d}^3(mv) = (1/m^4) F$.}:
\begin{equation}
   \label{eq:phase_space_bound_condition}
   F^{\mathrm{late}}_{\mathrm{coarse}} \leq F^{\mathrm{late}}_{\mathrm{fine}} \leq F^{\mathrm{prim.}}_{\mathrm{fine}} = m^4 f_{\mathrm{max}}\ ,
\end{equation}
where the first inequality is as a result of the explicit coarse-graining, and the second is the application of Liouville's theorem. Here $f_\mathrm{max}$ denotes the maximum of the primordial distribution function. 

Given that we are interested in strengthening the currently existing phase-space bounds on the fermionic dark matter mass, Eq.~\eqref{eq:phase_space_bound_condition} provides  the following three ways to achieve this (and of which in this work we will exploit the first and third options):

\begin{itemize}
    \item \textit{An improved coarse-graining ---} Since the phase-space density is inversely proportional to the volume in velocity space, a more accurate determination of the velocity dispersion at late times within the halo will increase the observed, coarse-grained phase-space density $F^{\mathrm{late}}_{\mathrm{coarse}}$ and thus make the bounds stronger. 
    \item \textit{Explicit tracking of the distribution function evolution ---}
    While Liouville's theorem states that the maximum of the primordial phase-space density cannot increase, it could be possible that $F^{\mathrm{late}}_{\mathrm{fine}}$ actually decreases due to some dynamical effects, such as mixing induced by merger events or randomisation of bulk motions, see e.g.~\cite{Peirani:2005kw, Piattella:2013cma} and references therein. N-body simulations of fermionic dark matter that track the evolution of the phase-space density have shown that typically this is indeed the case \citep{Shao:2012cg}. However, since this prediction is not completely quantitative as of yet, we use the maximally allowed theoretical value for $f_\mathrm{max}$ in order to be conservative.
    \item \textit{Better determination of the primordial distribution ---} Fermionic particles obey Pauli's principle, which imposes a stringent upper limit on the maximum of their distribution function. This provides the most conservative case for any given fermionic species. Different production scenarios, however, can cause this maximum to deviate from this number and therefore strengthen the bounds. In our work, we use the best possible determination of $F^{\mathrm{prim.}}_{\mathrm{fine}}$ from either theoretical considerations or simulations. In the event that it is not possible to accurately do this, one can always fall back to the Pauli principle. This is particularly relevant in deriving our bounds on resonantly produced sterile neutrinos in Sec.~\ref{sec:sterilenu}, where further discussion can be found.
\end{itemize}

\subsection{The Observed Coarse-grained Phase-space Density}\label{sec:coarsegrained}

At the simplest level, the breaking of the $\beta$-degeneracy in the Jeans analysis allows us to far better constrain the dark matter density profile $\rho(r)$ in the inner, most dense, regions of the dwarf. Indeed, as we will see later on, it is in this region that the observed phase-space density is highest. Now, we will describe two methods to obtain a coarse-grained phase-space density: \emph{(i)} based on maximal coarse-graining and \emph{(ii)} based on coarse-graining with physically motivated assumptions about the velocity distribution of the dark matter.

\vspace*{-0.7cm}

\subsubsection{Maximal Coarse-graining --- if you only know $\rho(r)$}\label{sec:maxcoarsegraining}
Consider the case that one only has access to the density profile of the dwarf galaxy $\rho(r)$. Then one can be sure that the magnitude of the velocity of the dark matter particles is less than the escape velocity $v_\mathrm{esc}$. We generalise the approach in~\cite{Boyarsky:2008ju} to take advantage of the full density profile $\rho(r)$, as opposed to just the total enclosed mass at the half-light radius. To do so, we consider a phase space volume that consists of a spherical shell at radius $r$ and a velocity volume given by: 
\begin{equation}
    \Gamma(r) =\frac{4}{3}\pi v_\mathrm{esc}(r)^3\ .
\end{equation}
The coarse-grained phase-space density is then:
\begin{equation}
    \label{eqn:finf}
    \overbar{F}_\mathrm{M}(r) =\frac{\rho(r)}{\Gamma(r)}=\frac{3 \rho(r)}{4 \pi v_{\mathrm{esc}}(r)^{3}}.
\end{equation}
We call this type of coarse-graining \emph{maximal coarse-graining}, as it covers the entirety of the velocity phase-space volume in which the particles reside. Using Eq.~\eqref{eq:phase_space_bound_condition}, we see that an application of Liouville's theorem allows us to set a bound via:
\begin{align}
    \label{eqn:finf_bound}
    \boxed{\overbar{F}_\mathrm{M}(r) = \frac{3 \rho(r)}{4 \pi v_{\mathrm{esc}}(r)^3} \leq m^4 f_{\mathrm{max}}\ .}
\end{align}
\noindent A bound obtained in this way is the most conservative, as it does not make any assumptions regarding the velocity distribution of the dark matter particles. In particular, this does not require any knowledge about the velocity anisotropy parameter $\beta_{\mathrm{dm}}$.

\vspace*{-0.5cm}

\subsubsection{Gaussian Coarse-graining --- using the velocity dispersion}\label{sec:gausscoarsegraining}

We can improve this estimate of the maximal coarse-grained phase-space density by considering the dynamics of the dark matter within the halo. In particular, if we solve the Jeans equation for the dark matter with a suitable prior on the $\beta_{\mathrm{dm}}$ profile (see Sec.~\ref{subsec:DM_Jeans}), we can derive the radial and tangential velocity dispersions of the particles.

If we combine this with an assumption about the form of the dark matter velocity distribution, we do not need to coarse-grain the velocity phase-space volume. Instead we can analytically find the maximum phase-space density and significantly improve our previous estimate. A commonly used choice for the velocity distribution of dark matter inside a virialised halo is that of a multivariate Gaussian. Such a distribution provides a reasonable description of the dynamics as motivated by N-body simulations~\citep{Vogelsberger:2008qb}. Denoting the radial and tangential velocity dispersions as $\sigma_r$ and $\sigma_t$ respectively, the coarse-grained phase-space density is given by:
\begin{equation}
    \overbar{F}_\mathrm{G}(r, \mathbf{v}) = \frac{\rho(r)}{(2\pi)^{3/2} \sigma_r \sigma_t^2} \exp \left[-\frac{1}{2}\left(\frac{v_r^2}{\sigma_r^2} + \frac{v_\theta^2}{\sigma_t^2} + \frac{v_\phi^2}{\sigma_t^2}\right)\right]\ ,
\end{equation}
where $\mathbf{v} = (v_r,v_\theta,v_\phi)$ is the velocity vector. We call this procedure \emph{Gaussian coarse-graining} to distinguish it from the approach in the previous section. It can easily be seen that the maximum of this distribution occurs when $v_r = v_\theta = v_\phi = 0$, which reads\footnote{It is worth noting here that it has been common in the literature to use the expression $Q = \rho / (3^{3/2}\sigma^3)$ as an estimator of the coarse-grained phase-space density in Liouville's theorem. We emphasize that this is not a correct procedure, as Liouville's theorem explicitly states that the maximum of the distribution function cannot increase, not its average. Moreover, a number of works, e.g. \cite{Boyarsky:2008ju}; \cite{Shao:2012cg}, have pointed out that whilst this quantity has the correct dimensions, it is a significant \emph{overestimate} of the true phase-space density. Indeed, we see from Eq.~\eqref{eqn:fg} that the value of $Q$ is about 3 times larger. The intuition behind this is that the expression $Q = \rho / (3^{3/2}\sigma^3)$ does not take into account the fact that a large proportion of the dark matter particles will have a velocity \emph{larger} than $\sqrt{3}\sigma$, leading to an overestimate. With this in mind, we follow the approach in Eq.~\eqref{eqn:fg} to derive the bounds in this work.}:
\begin{equation}\label{eqn:fg}
    \overbar{F}_\mathrm{G}^\mathrm{max}(r) =  \frac{\rho(r)}{(2\pi)^{3/2}\sigma_r(r)\sigma_t^2(r)}\ .
\end{equation}

\begin{figure}
    \centering
    \includegraphics[width=\linewidth]{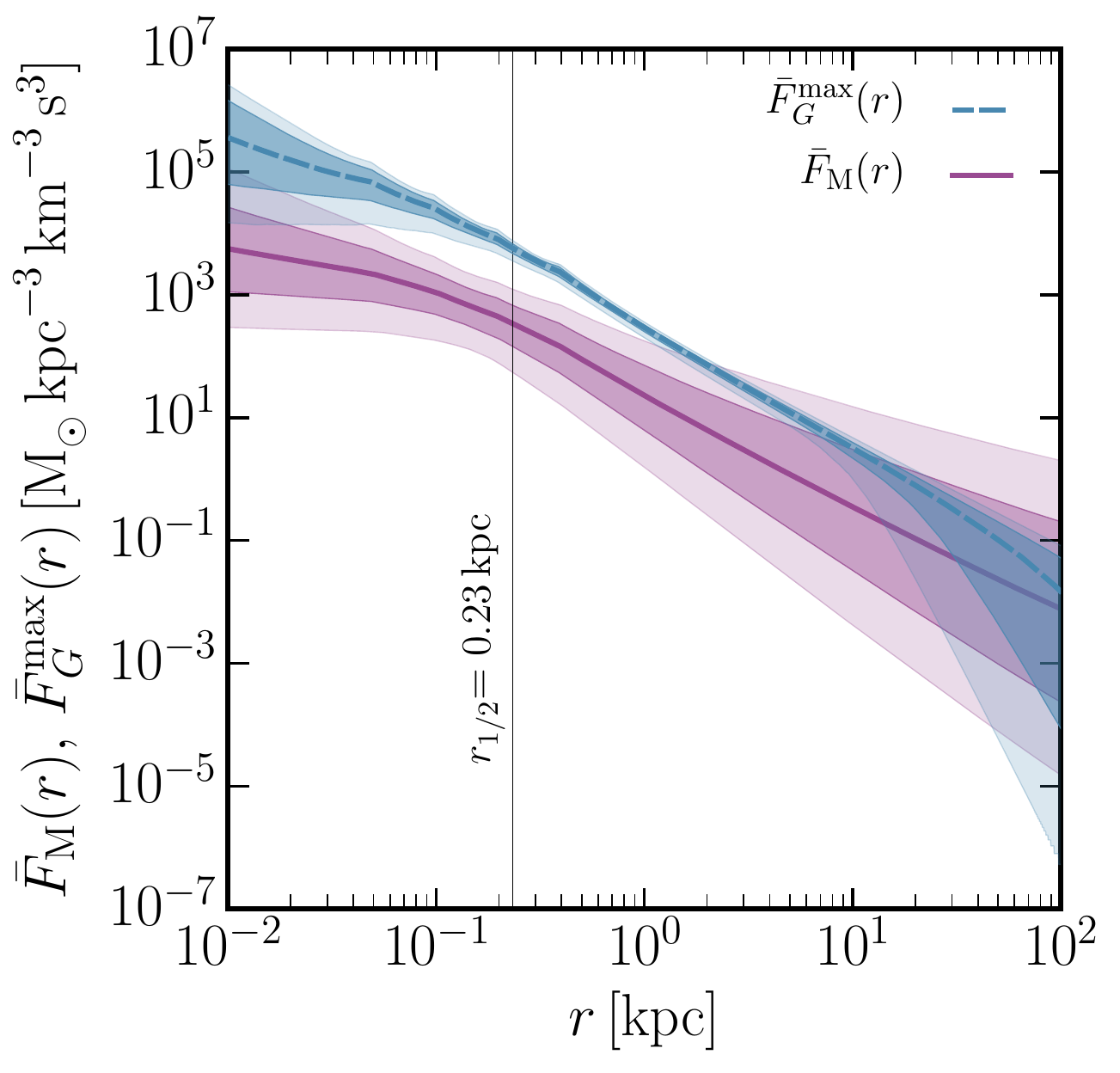}
    \caption{Comparison between the coarse-grained phase-space densities $\bar{F}_\mathrm{M}(r)$ and $\bar{F}_G^{\mathrm{max}}(r)$ for the dwarf galaxy Leo II. These are computed as explained in Eqs.~\eqref{eqn:finf} and~\eqref{eqn:fg}, respectively. The solid/dashed line refers to the average value of the quantity whilst the darker and lighter shaded regions are the 68\% and 95\% CL regions respectively. The value $r_{1/2}$ refers to the 3D projected half-light radius of the galaxy.
    }
    \label{fig:Fbar_Leo2}
\end{figure}

\noindent In a similar fashion to the case of maximal coarse-graining, this expression can again be used in Eq.~\eqref{eq:phase_space_bound_condition} to obtain lower bounds on the dark matter mass via:
\begin{equation}
    \label{eq:fg_bound}
    \boxed{\overbar{F}_\mathrm{G}^\mathrm{max}(r) =  \frac{\rho(r)}{(2\pi)^{3/2}\sigma_r(r)\sigma_t^2(r)} \leq m^4 f_{\mathrm{max}}\ .}
\end{equation}
There is a slight caveat in using this expression to derive a bound, in that for certain masses Eq.~\eqref{eqn:fg} can exceed the maximum phase-space density allowed by the Pauli principle. We account for this potential issue by imposing the phase-space density of a degenerate gas as an upper bound on the inferred phase-space density.

Fig.~\ref{fig:Fbar_Leo2} shows the coarse-grained phase-space densities using the two different methods discussed above. In Sec.~\ref{sec:Results}, we will present bounds on the dark matter mass using both the \emph{maximal} and \emph{Gaussian} coarse-graining methods.

\section{Benchmark Particle Physics Models}
\label{sec:benchmark_models}
As discussed in the previous section, and as is evident from Eq.~\eqref{eq:phase_space_bound_condition}, in order to set bounds on the dark matter mass we need to find the maximum value of the primordial distribution function of a fermionic dark matter particle. Clearly, this depends on the production mechanism of the species and as such we examine a number of benchmark models here.

\subsection{Relativistically decoupled thermal fermions}
Assuming that the halo consists of a relativistically decoupled thermal relic, the primordial distribution function is:
\begin{align}
    \label{eq:f_FD}
    f^\mathrm{FD}(p) = \frac{g}{(2\pi)^3}\frac{1}{e^{p/T_\mathrm{dec}}+1} \quad \longrightarrow \quad f^{\mathrm{FD}}_\mathrm{max} = \frac{g}{2(2\pi)^3}\ ,
\end{align}
where $T_{\mathrm{dec}}$ is the decoupling temperature and $g$ is the number of internal degrees of freedom. We note that $f^{\mathrm{FD}}_\mathrm{max}$ is independent of $T_\mathrm{dec}$ and fermion mass $m_\mathrm{FD}$. Thus, we can place model-independent bounds on $m_\mathrm{FD}$ provided that the halo is formed entirely out of species described by Eq.~\eqref{eq:f_FD} in the early Universe\footnote{Note that requiring this particle to represent the dark matter of the Universe will yield a relationship between its temperature today and its mass.}. Within a cosmological setting, such a particle does not represent the most realistic dark matter candidate. As we will see later on, however, this benchmark scenario will be useful when deriving bounds for other models.

\subsection{Sterile Neutrinos}\label{sec:sterilenu}
\label{subsubsec:sterile_neutrinos}
Sterile neutrinos are right-handed companions to the Standard Model active neutrinos and have recently seen increasing interest as a dark matter candidate (see e.g.~\cite{Adhikari:2016bei,Boyarsky:2018tvu} for reviews on this topic). Fueled by the discovery of an as-of-yet unidentified emission line in the X-ray spectrum of galaxies and galaxy clusters~\citep{Bulbul:2014sua,Boyarsky:2014jta}, ongoing efforts have been made to constrain its parameter space in the keV region from an experimental~\citep{Mertens:2014nha}, astrophysical, as well as a cosmological perspective~\citep{Boyarsky:2009ix, Boyarsky:2012rt}.

Phase-space constraints on keV sterile neutrinos have been derived before in e.g.~\cite{Gorbunov:2008ka, Boyarsky:2008ju, Horiuchi:2013noa, Wang:2017hof} and, together with those from UV luminosity function~\citep{Menci:2016eww, Menci:2017nsr}, Milky Way satellite counts~\citep{Cherry:2017dwu} and Ly-$\alpha$ observations~\citep{Garzilli:2018jqh}, have been found to be complementary to those from X-ray limits~\citep{Boyarsky:2018tvu}. Each of the bounds mentioned above have a number of astrophysical uncertainties associated with them. It is therefore important to provide robust lower limits so as to tighten the constraints in this region of parameter space.

As we emphasised before, in order to derive a bound, we must also consider their cosmological production. In this work, only minimal models that involve sterile neutrinos are considered. We cover two of the most prominent ones in detail: \emph{(i)} non-resonant and \emph{(ii)} resonant production. Moreover, we also discuss alternative production modes of sterile neutrinos, such as through the decay of scalar particles in the early Universe.

\subsubsection{Non-resonant Production} 
Sterile neutrinos can be produced via oscillations with active neutrino species in the early Universe. In the absence of a primordial lepton asymmetry, this proceeds via the well-known Dodelson-Widrow mechanism~\citep{Dodelson:1993je} and is typically termed \emph{non-resonant} production. In this same reference it has been shown that at low temperatures the non-resonant sterile neutrino distribution is roughly proportional to that of the active neutrino. This has been further established by a comparison between the thermal-like approach and that obtained by solving a Boltzmann equation, e.g.~\citep{Abazajian:2005gj}. For masses relevant to this work, an agreement within ${\sim}20\%$ was found in the low momentum region where the maximum of the distribution is attained\footnote{In terms of bounds on the mass of the sterile neutrino, this will induce an error of ${\sim}6\%$, which is subdominant compared to those from the Jeans analysis.}. As such, we follow~\cite{Boyarsky:2008ju}, where we assume that the primordial sterile neutrino distribution is approximately thermal\footnote{We have tried to use the public codes \texttt{sterile-dm}~\citep{Venumadhav:2015pla} and \texttt{resonance-dm}~\citep{Ghiglieri:2015jua} to validate this assumption, but found that both codes are not able to give reliable estimates for the maximum of the distribution function (which occurs at low momenta) in this configuration. However, both codes are able to reproduce the correct relic abundance and distribution function for non-resonant sterile neutrinos at the higher momentum range ($p/T \gtrsim 0.1$).} with a normalisation constant $\mathcal{N}$ which ensures that the particle constitutes the entirety of the dark matter. The maximum of the distribution is thus:
\begin{align}
    \label{eq:f_NRP}
    f^\mathrm{NRP}(p) = \frac{g\mathcal{N}}{(2\pi)^3}\frac{1}{e^{p/T_\mathrm{dec}}+1} \quad \longrightarrow \quad f^{\mathrm{NRP}}_\mathrm{max} = \frac{93\,\mathrm{eV}}{m}\frac{\omega_\mathrm{dm}}{2(2\pi)^3}\ ,
\end{align}
where $\omega_\mathrm{dm} = 0.12$~\citep{Aghanim:2018eyx}.

\subsubsection{Resonant Production} 
\label{subsubsec:rp_sterile}
In the presence of a primordial lepton asymmetry, the production of sterile neutrinos proceeds through an MSW-like effect. In this case, the lepton asymmetry in the Standard Model neutrino sector gets transferred to the sterile neutrino sector and leads to resonant peaks in their distribution function. This is also known as the Shi-Fuller mechanism~\citep{Shi:1998km}. 

The existence of such resonance peaks means that the exact shape of the sterile distribution function in this mechanism can be computed only numerically. Currently there are two codes available that are able to do this: \texttt{sterile-dm}~\citep{Venumadhav:2015pla} and \texttt{resonance-dm}\footnote{Note that this code defines the sterile neutrino as a Dirac particle with distribution $f^+ = (f_\mathrm{s} + \bar{f}_\mathrm{s})/2$, where $f_\mathrm{s}$ and $\bar{f}_\mathrm{s}$ denote the true distributions of the sterile neutrino and its charge conjugate. The helicity term $f^- = (f_\mathrm{s} - \bar{f}_\mathrm{s})/2$ is not included in the equation of motion for $f^+$~\citep{Ghiglieri:2015jua}. However, as long as $f^+$ (and thus $f^-$ as $|f^-| \leq f^+$) are both much smaller than the Fermi-Dirac distribution $f^\mathrm{FD}$, the contribution of $f^-$ to the equation of motion for $f^+$ remains small~\citep{Ghiglieri:2019kbw, Ghiglieri:2020ulj}. This means that for relatively small mixing angles and large lepton asymmetries (when $f_\mathrm{FD} \gg f_\mathrm{s} \gg \bar{f}_\mathrm{s}$) the sterile neutrino distribution can be obtained via $f_\mathrm{s} \approx 2f^+$. We have confirmed this also by directly comparing with \texttt{sterile-dm}, which outputs $f_\mathrm{s}$}~\citep{Ghiglieri:2015jua}. The two codes have different regions of validity in the $(m_{\mathrm{s}}, \sin^2 2\theta)$ parameter space due to the different assumptions that are made in the numerical implementations. In what follows, we define the primordial lepton asymmetry as:
\begin{align}\label{eq:L6definition}
    L_6 = 10^6\, \frac{n_{\nu_\mu} - n_{\bar{\nu}_{\mu}}+n_{\mu^-} - n_{\mu^+}}{s}\ ,
\end{align}
where $n_i$ is the number density of particle species $i$ and $s$ is the entropy density of the Universe. Any asymmetries in the other flavours are set equal to zero. Since \texttt{sterile-dm} only operates with a lepton asymmetry in the muon flavour sector, we made this choice to easily compare between the two codes\footnote{At temperatures relevant for the production of sterile neutrino dark matter, $T \sim \mathcal{O}(100)\,\mathrm{MeV}$, electrons are relativistic and therefore abundant in the plasma, while the abundance of charged tau leptons is heavily suppressed. As such, we expect that in the case of electron mixing our results will not change significantly. On the other hand, for the tau mixing scenario, the majority of the lepton asymmetry at $T\sim 100\,\mathrm{MeV}$ will be in the tau neutrino sector~\cite{Venumadhav:2015pla}, which could strongly enhance the production of sterile neutrino dark matter for a fixed value of $L_6$. This may weaken both the BBN and phase-space bounds shown in Fig.~\ref{fig:RP_sterile}.}.

We have found that with the following procedure, we can obtain robust bounds on the sterile neutrino mass:
\begin{itemize}
    \item \emph{Large Mixing Angles --- } At large mixing angles ($\sin^2 2\theta \sim 10^{-9}-10^{-7}$), the lepton asymmetry required to obtain the correct relic abundance is relatively small ($L_6 \lesssim 100$ at beginning of production). Importantly, in this regime the sterile neutrino distribution function approaches that of a Fermi-Dirac distribution $f_s(p) \simeq f^\mathrm{FD}(p)$ at low momenta. The code \texttt{resonance-dm} breaks down in this regime, as it implicitly assumes that that the sterile distribution is much smaller than the Fermi-Dirac distribution. As such, we cannot use it to model the behaviour in this area of parameter space. Nonetheless, we can conservatively estimate that the maximum of the distribution does not exceed that of a Fermi-Dirac distribution, $f_{\mathrm{max}}^{\mathrm{RP}} \leq g / (2(2\pi)^3)$. This is because the sterile neutrinos typically try to equilibrate with the active neutrinos, which follow a thermal distribution at temperatures of around $100\,\mathrm{MeV}$. At large masses $m_s \gtrsim 3 \, \mathrm{keV}$ we have explicitly confirmed this with \texttt{sterile-dm}.\vspace{6pt}
    \item  \emph{Intermediate Mixing Angles --- } With the restriction that the sterile neutrinos constitute the entirety of the dark matter, the lepton asymmetry in this regime of mixing angles ($\sin^2 2\theta \sim 10^{-12}-10^{-9}$) is considerably larger ($100 \lesssim L_6 \lesssim 2500$ initially). For masses of a few keV and mixing angles in this range, \texttt{sterile-dm} consistently underproduces the sterile neutrino dark matter. The reason for this behaviour is unknown. Within this regime, however, \texttt{resonance-dm} is able to return the correct relic abundance. Moreover, we have explicitly checked that the required assumptions regarding the hierarchy of distribution functions ($f^\mathrm{FD} \gg f_s \gg \overbar{f}_s$) hold in this regime. As such, we can use \texttt{resonance-dm} to compute the distribution functions and derive the most up-to-date bounds on this region of parameter space\footnote{There is a final technicality that the code does not output the distribution at momentum $p = 0$, where we expect the maximum to be. However, we have computed the abundance of those dark matter particles with momenta below the lowest momentum point given by the code ($p\sim0.003\,\mathrm{MeV}$) and found that they only contribute a tiny fraction (${\sim}10^{-7}$) to the total abundance. Therefore, this approach is valid under the assumption that there are no processes that can significantly increase the abundance of low-momentum ($p\lesssim0.003\,\mathrm{MeV}$) dark matter particles.
    }.\vspace{6pt}
    \item  \emph{Small Mixing Angles --- } At very small mixing angles ($\sin^2 2\theta \lesssim 10^{-12}$), a large primordial lepton asymmetry is required ($L_6 \gtrsim  2500$) in order to reproduce the correct relic abundance of the dark matter. Such large primordial lepton asymmetries are disfavoured by successful BBN that requires $L_6(T \sim 1 \, \mathrm{MeV}) \lesssim 2500$ at 95\% CL. This value for the lepton asymmetry was obtained using the relations in~\cite{Pitrou:2018cgg} and corresponds to an error in the primordial helium abundance of ${\sim}4.5\%$, which is a conservative estimate based on a comparison of the studies in~\cite{Izotov:2014fga,Aver:2015iza,Peimbert:2016bdg,Fern_ndez_2018,Valerdi:2019beb} and takes into account potential systematic errors in the primordial helium determination. Note that using a smaller error in the helium abundance will not change the BBN bound significantly, as the lepton asymmetry quickly decreases with increasing mixing angle.
    Therefore, sterile neutrino dark matter with these small mixing angles is disfavoured by this consideration. In addition, one should note that the codes \texttt{sterile-dm} and \texttt{resonance-dm} expand the equations of motion to leading order in the chemical potential, and thus fail in this regime. Therefore, we do not model these regions of parameter space, although we are able to extend the bound from larger mixing angles vertically downwards to obtain a conservative constraint (as the maximum of the distribution continues to decrease with smaller mixing angles). Of course, this region is also covered by the independent bound from BBN.
\end{itemize}

\begin{figure}
    \vspace{-0.2cm}
    \centering
    \includegraphics[width=\linewidth]{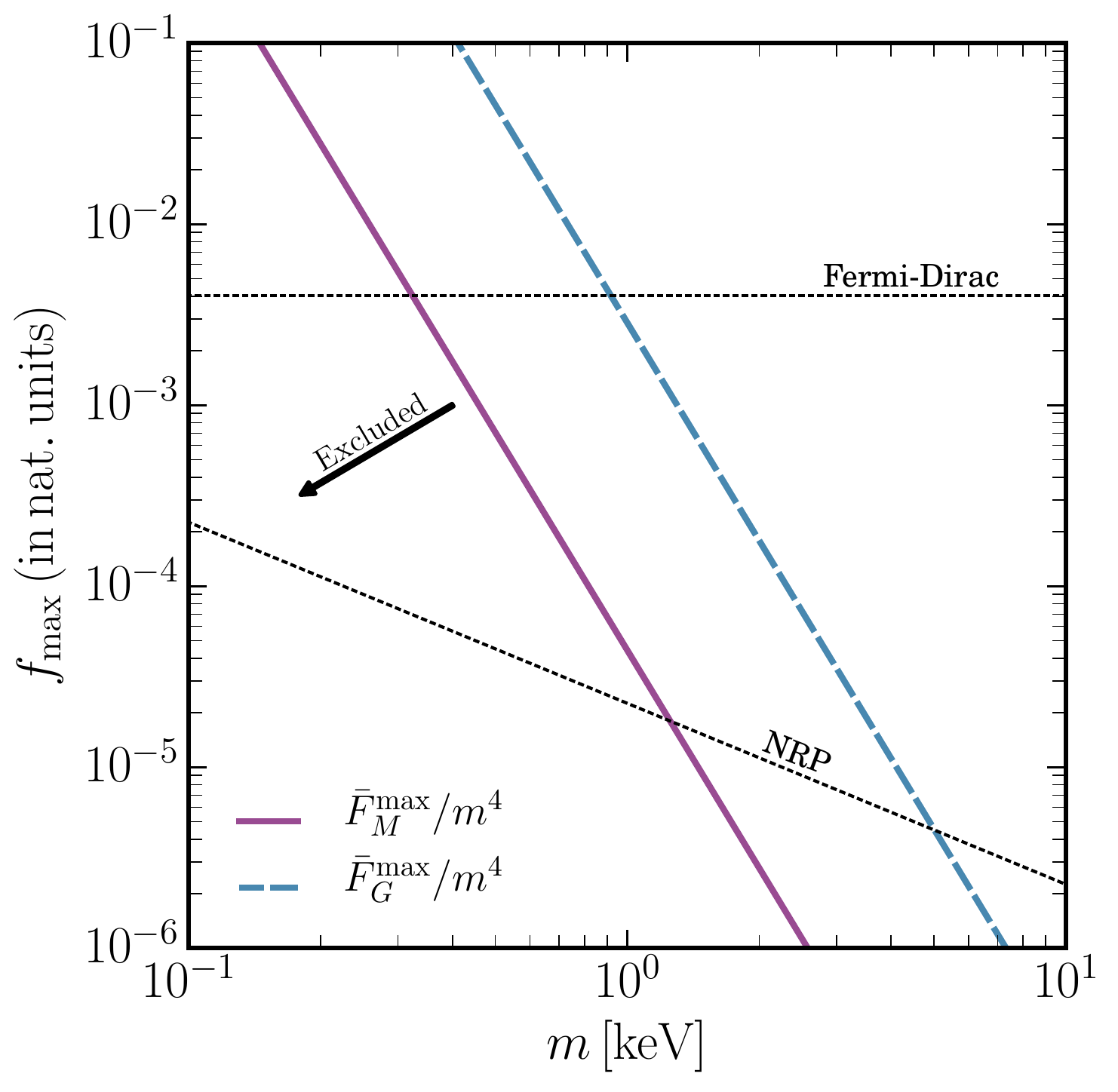}
    \caption{
    Bounds on the maximum of the primordial distribution function of fermionic dark matter $f_\mathrm{max}$ as a function of its mass. The purple and blue curves are the coarse-grained phase-space densities obtained through maximal and Gaussian coarse-graining respectively (see Sec.~\ref{sec:coarsegrained} for details). 
    This bound is based on Eq.~\eqref{eq:phase_space_bound_condition}, where a mass range is excluded if $f_\mathrm{max}$ is below the solid/dashed line. The benchmark models from Sec.~\ref{sec:benchmark_models} are added here as illustration (e.g. Eq.~\eqref{eq:f_FD} for a relativistically decoupled thermal particle), but the solid purple and dashed blue curves can be used for any fermionic dark matter model to infer phase-space constraints on its parameter space.}
    \label{fig:fmax_vs_mass}
\end{figure}

\subsubsection{Alternative Production Scenarios}

The most studied production mechanisms of keV-scale sterile neutrinos in the early Universe are scatterings and oscillations (both non-resonant and resonant) with active neutrinos. There are, however, several other possibilities for sterile neutrino dark matter production. A sample of alternative possibilities are: sterile neutrino production from the decays of heavier particles, including singlet scalars~\citep{Shaposhnikov:2006xi,Kusenko:2006rh,Konig:2016dzg}, Dirac fermions~\citep{Asaka:2006ek,Abada:2014zra} or charged scalars~\citep{Frigerio:2014ifa}, and sterile neutrino production assisted by neutrinophilic scalars~\citep{deGouvea:2019phk} or gauge bosons~\citep{Nemevsek:2012cd}. Importantly, regardless of the production mechanism, the very same procedure to set bounds on the sterile neutrino mass described in Sec.~\ref{sec:HowToObtainBounds} applies. Although it is beyond the scope of this paper to set constraints on each of these production mechanisms, we note that our procedure typically yields a bound $m_\mathrm{s} \gtrsim \mathcal{O}(1)\,\text{keV}$ for sterile neutrinos produced via these alternative mechanisms. As an illustrative example of how to use our derived bounds, one can consider the recent model of~\cite{deGouvea:2019phk}. Taking the benchmark scenario denoted as B in that reference, the maximum of the distribution is $f_\mathrm{max}
^{\rm B} \simeq 2\times 10^{-3}$ for $m_s = 7.1\,\text{keV}$. From our Fig.~\ref{fig:fmax_vs_mass} we can therefore see that this point is unconstrained by our method. On the other hand, lower masses will be constrained by our approach.

\section{Results}
\label{sec:Results}
This section is devoted to the application of the tools developed in Sec.~\ref{sec:HowToObtainBounds} on the benchmark models discussed in Sec.~\ref{sec:benchmark_models}. Throughout the text, we report phase-space bounds on the fermionic dark matter mass based on data from Leo II, where the strongest constraints are found. We summarise the bounds\footnote{A full set of tabulated bounds can be found on the \href{https://github.com/james-alvey-42/FermionDSph}{\texttt{GitHub}} page.} for the other dwarfs in Fig.~\ref{fig:bounds_full_set}. All results are shown for fermions with $g = 2$ degrees of freedom.

\begin{table}
    \centering
{\def\arraystretch{1.35}
    \begin{tabular}{c|c|c|c|c}
        \hline\hline
         \multirow{2}{*}{\textbf{Model}} & \multicolumn{2}{c}{\textbf{Max. coarse-gr.}} & \multicolumn{2}{c}{\textbf{Gauss. coarse-gr.}}\\
         & $1\sigma$ & $2\sigma$ & $1\sigma$& $2\sigma$ \\
        \hline\hline
        \textbf{Thermal FD}& $0.32^{+0.15}_{-0.11}$ & $0.32^{+0.36}_{-0.17}$ & $0.92^{+0.38}_{-0.33}$ & $0.92^{+0.59}_{-0.50}$\\
        \hline
        \textbf{NRP sterile}&
        $1.25^{+0.85}_{-0.52}$& $1.25^{+2.16}_{-0.78}$ &
        $5.02^{+2.98}_{-2.22}$ & $5.02^{+4.71}_{-3.29}$ \\
        \hline
        \textbf{RP sterile}& \multicolumn{4}{c}{see Fig.~\ref{fig:RP_sterile}} \\
        \hline\hline
    \end{tabular}
}
    \caption{Phase-space bounds on the fermionic dark matter mass (in keV) for the benchmark models considered in this work using data from the Leo II dwarf galaxy. These constraints are based on Liouville's theorem (`Max. coarse-gr' from  Eq.~\eqref{eqn:finf_bound} and `Gauss. coarse-gr' from Eq.~\eqref{eq:fg_bound}).}
    \label{tab:Liouville_bounds}
\end{table}

\subsection{Bounds from Pauli's principle}
\label{subsec:bounds_pauli}
The density and escape velocity profiles for Leo II (see also Sec.~\ref{sec:JeansAnalysis}), are plugged in  Eq.~\eqref{eq:pauli_bound_condition} to obtain the most robust, model-independent phase-space constraint. The strength of the bound depends on the distance from the center, where we find that it becomes stronger towards the center of the dwarf. Hence, we report it at this innermost region. We find that the mass of the fermionic dark matter should satisfy:
\begin{align}
\label{eq:m_Pauli_result}
m_{\mathrm{deg}} \geq
\begin{cases}
0.27^{+0.13}_{-0.09}\,\mathrm{keV} & (1\sigma)\\
0.27^{+0.30}_{-0.14}\,\mathrm{keV} & (2\sigma)
\end{cases}\ .
\end{align}

\subsection{Bounds from Liouville's theorem}
\label{subsec:bounds_liouville}
We start by computing the coarse-grained phase-space densities in Eqs.~\eqref{eqn:finf} and \eqref{eqn:fg}.
Next, we apply the conditions in Eqs.~\eqref{eqn:finf_bound} and~\eqref{eq:fg_bound} to obtain the constraints for the benchmark models considered here. We summarise the results in Table~\ref{tab:Liouville_bounds}.

In addition, Eqs.~\eqref{eqn:finf_bound} and~\eqref{eq:fg_bound} allow us to directly set constraints on the maximum of the primordial distribution function $f_\mathrm{max}$, regardless of which model for the fermionic dark matter is considered. Therefore, by mapping the bound on $f_\mathrm{max}$ to the parameters of a specific particle physics model, phase-space bounds can be readily obtained. The bound on $f_\mathrm{max}$ as a function of the dark matter mass is shown in Fig.~\ref{fig:fmax_vs_mass}, where data from Leo II is used.

\begin{figure}
    \centering
    \includegraphics[width=\linewidth]{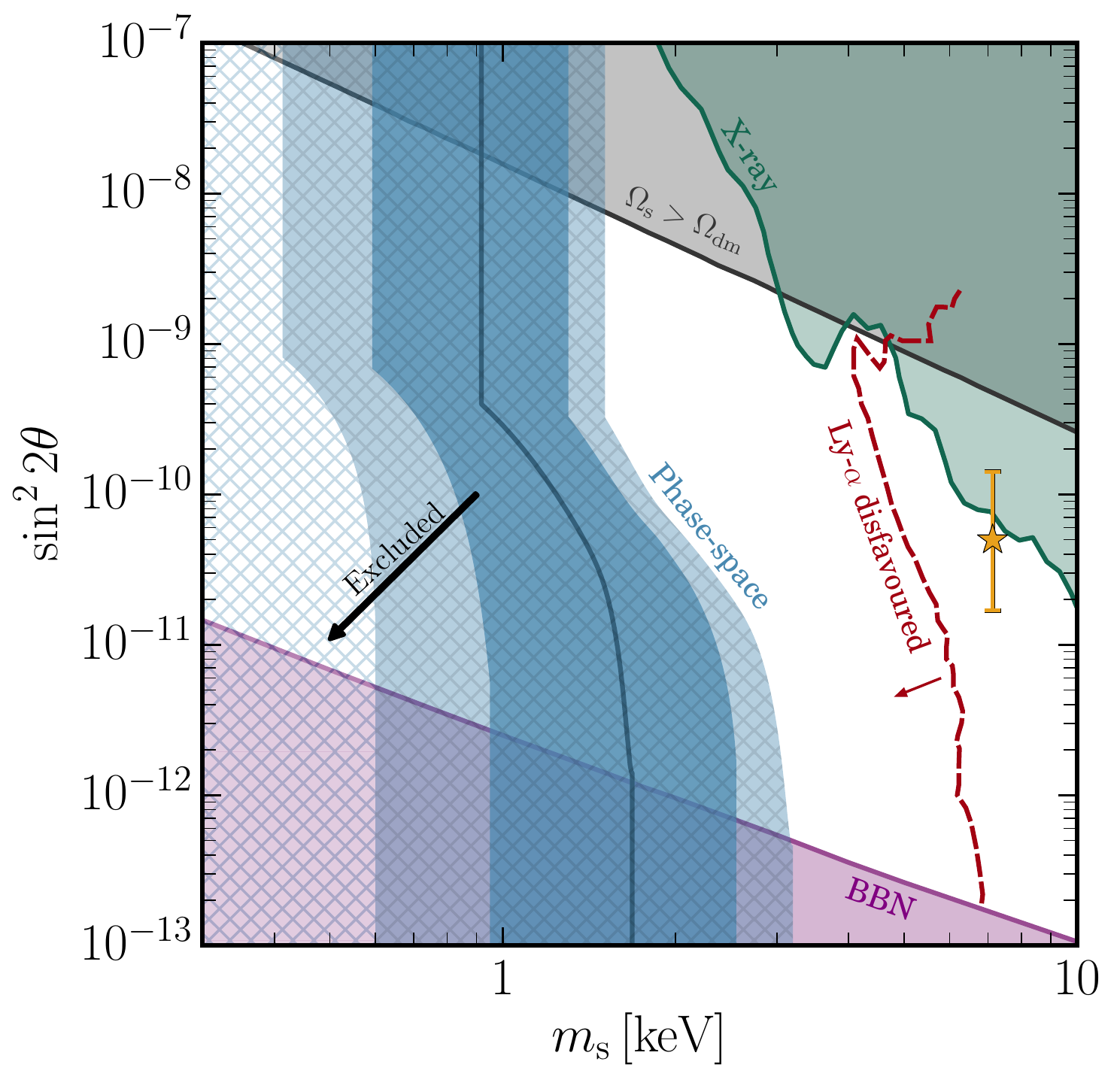}
    \vspace{-6pt}
    \caption{Bounds on resonantly produced sterile neutrinos. The blue, hatched area is the phase-space constraint based on Gaussian coarse-graining. The dark blue line indicates the central value of the bound, below which masses are excluded (as indicated by the arrow). The dark (light) blue regions are the $\pm1\sigma$ ($\pm2\sigma$) confidence intervals. The purple region represents the BBN constraint on the primordial lepton asymmetry, see Sec.~\ref{subsubsec:rp_sterile} for details. The black line depicts where sterile neutrinos are non-resonantly created and above which the dark matter is overproduced. Complementary constraints from
    X-ray observations~\citep{Boyarsky:2018tvu}
    and indicative sensitivity of the
    Lyman-$\alpha$ forest method~\citep{Baur:2017stq} are also included, see Sec.~\ref{subsec:bounds_liouville} for comments. The orange star with error bars denotes the sterile neutrino interpretation of the tentative signal recently observed in X-ray data~\citep{Boyarsky:2014jta}.}
    \label{fig:RP_sterile}
    \vspace{-7pt}
\end{figure}

Finally, we report phase-space bounds on resonantly produced sterile neutrinos in Fig.~\ref{fig:RP_sterile}. These bounds are obtained using the method described in Sec.~\ref{subsubsec:rp_sterile}. Note that the phase-space constraints in this figure are based on Gaussian coarse-graining, while for maximal coarse-graining they are a factor ${\sim}2.8$ weaker. Our results are also put in a wider context by including complementary bounds from BBN (Sec.~\ref{subsubsec:rp_sterile}) and X-ray and Lyman-$\alpha$ studies. 

We would like to stress here that, although the Lyman-$\alpha$ method is potentially more powerful in constraining resonantly produced sterile neutrinos than the phase-space one described here, it is also much more indirect and subject to very hard-to-control physical uncertainties. In particular, a major uncertainty is the thermal history of hydrogen, which could affect Lyman-$\alpha$ absorption spectra in at least two different ways: \emph{(i)} by the Doppler widening of the lines which destroys correlations at small scales, and \emph{(ii)} by pressure effects that physically prevent hydrogen from following dark matter at small scales. For some thermal histories, these effects can explain the small scale cut-off observed in the Lyman-$\alpha$ forest power spectrum at redshifts $4.5 < z < 5.5$. For other histories, the data becomes inconsistent with $\Lambda$CDM and requires, for example, that the dark matter be warm (see~\cite{Garzilli:2015iwa}). This makes high-resolution data to some extent inconclusive and the bound that can be derived from such data quite weak~\citep{Garzilli:2019qki}. Experiments such as SDSS/BOSS probe the power spectrum at larger scales, where no such cut-off is observed. In this case, the high statistics of sources available in these surveys leads to statistical error bars that are quite small, and hence, the formal sensitivity of the method is rather high. On the other hand, it is unclear to what extent the predictions for the power spectrum at these scales are sensitive to the unknowns in the thermal history and whether the subsequent physical/systematic uncertainties are at the same level as the statistical ones. Arguably, this is not the case yet, and ensures that although the Lyman-$\alpha$ method is potentially very constraining, it is significantly less robust and conservative than the phase-space approach described here.

\begin{figure*}
    \centering
    \includegraphics[width=1.\linewidth]{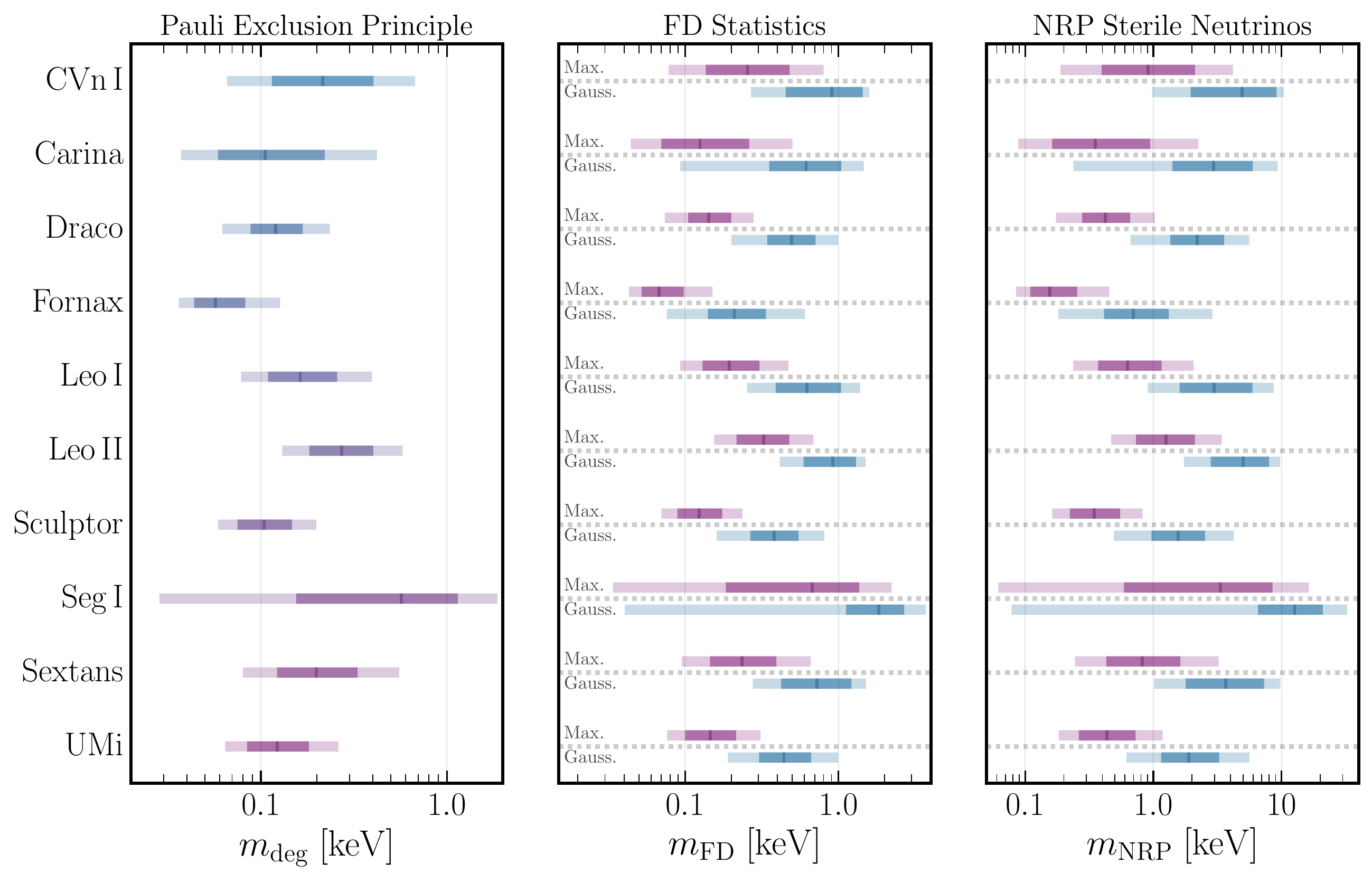}
    \vspace{-0.2cm}
    \caption{Bounds on the mass of fermionic dark matter for the full set of dwarfs considered in this work. These are computed using the Pauli exclusion principle (\emph{left}) and Liouville's theorem for relativistically decoupled thermal fermions (\emph{middle}) and non-resonantly produced sterile neutrinos (\emph{right}). The $1\sigma$ and $2\sigma$ constraints are represented by the darker and lighter colours respectively. The dark vertical line inside the $1\sigma$ constraint is the central value. For each dwarf, the bounds above the dotted lines are obtained using maximal coarse-graining, while those below are obtained using Gaussian coarse-graining.}
    \label{fig:bounds_full_set}
\end{figure*}

\subsection{Comparison with previous studies}
 A number of studies in the literature have obtained phase-space bounds on fermionic dark matter. These include~\cite{Dalcanton:2000hn, DiPaolo:2017geq, Savchenko:2019qnn}, where constraints on degenerate and/or relativistically decoupled thermal fermionic dark matter particles were obtained. Other works, e.g.~\cite{Gorbunov:2008ka, Horiuchi:2013noa, Wang:2017hof}, have applied phase-space considerations to sterile neutrino dark matter. 
 
 In this section, our main comparison is done with~\cite{Boyarsky:2008ju}, as we follow a somewhat similar methodology. Nonetheless, we will note throughout the text that some of our conclusions also apply to the methods utilised in the other references. For convenience, we split the comparison with~\cite{Boyarsky:2008ju} into a discussion around the calculation of the fine- and coarse-grained phase-space densities, followed by a comment on the way the bounds are subsequently obtained.

\subsubsection{Fine-grained phase-space density}\label{sec:finegrainedcomparison}
We consider the same benchmark particle physics models as~\cite{Boyarsky:2008ju}. The main difference lies in the prediction of the maximum phase-space distribution of resonantly produced sterile neutrinos. This reference uses distribution functions that are obtained with a code based on the work in~\cite{Laine:2008pg}. There are two factors to consider here: firstly, their distribution functions stop at a momentum $p\sim 0.03\,\mathrm{MeV}$, while ours extend down to $p\sim 0.003\,\mathrm{MeV}$. This roughly introduces a factor of 2 difference in the value of $f_\mathrm{max}$, since the distribution function is still increasing in this momentum range. Secondly, as with \texttt{resonance-dm}, the distributions they used are defined as $f_+ = \frac{1}{2}(f_\mathrm{s}+\bar{f_\mathrm{s}})$, not $f_\mathrm{s}$. For lepton asymmetries $L_6 \gtrsim 10$, this becomes $f_+ \sim \frac{1}{2}f_\mathrm{s}$, introducing an additional factor of 2. \cite{Boyarsky:2008ju} did not account for this multiplicative factor when they obtained $f_\mathrm{s}$. We have confirmed this by directly comparing their distributions with those obtained in \texttt{resonance-dm}. Together, these differences result in a factor ${\sim}4$ decrease of their $f_\mathrm{max}$ compared to ours and thus a factor $4^{1/4} \approx 1.4$ overestimation in their mass bounds. 

Other references that considered resonant sterile neutrinos either used \texttt{sterile-dm}~\citep{Wang:2017hof}, which we found to be not valid at low masses (see discussion in Sec.~\ref{subsubsec:rp_sterile}), or again used a rough version of the results from~\cite{Laine:2008pg} (see~\cite{Gorbunov:2008ka}). 

\subsubsection{Coarse-grained phase-space density}
The study in~\cite{Boyarsky:2008ju} assumed an isothermal distribution of stars and a constant dark matter density within the half-light radius of each dwarf. Moreover, the velocity distribution of the dark matter was taken to be isotropic, such that its escape velocity is given by $v_\mathrm{esc} = \sqrt{6}\sigma_*$, where $\sigma_*$ is the velocity dispersion of the stars. The latter quantity was then obtained from several studies available in the literature. As a result, the phase-space density only depends on $\sigma_*$ and the half-light radius.

Our work builds upon the above approach in two ways. Firstly, an improved Jeans analysis is utilised to obtain the density profile of the dark matter at radii smaller than the half-light radius. In particular, the \texttt{GravSphere} code simultaneously fits both surface density and velocity dispersion profiles, using photometric and stellar-kinematic data detailed in~\cite{2019MNRAS.484.1401R}. Secondly, another Jeans equation is solved for the dark matter itself to obtain the velocity dispersion. This allows us to relax the assumption that $\sigma_\mathrm{dm} = \sigma_*$, by imposing a prior on $\beta_\mathrm{dm}$ that is motivated by N-body simulations and physical intuition (see Sec.~\ref{subsec:DM_Jeans}). In this way, a more accurate determination of the dark matter phase-space density inside dwarf galaxies can be obtained, as described in Sec.~\ref{sec:HowToObtainBounds}.

The main bounds reported in~\cite{Boyarsky:2008ju} are obtained with Leo IV, which is not included in our set of dwarf galaxies. Therefore, as a concrete example, we compare our results for the galaxy Leo II. If we consider the maximal coarse-grained phase-space density evaluated at the half-light radius, then we find that the value obtained in our analysis is a factor ${\sim}10$ smaller than in this reference. The reason behind this is that their escape velocity ($v_\mathrm{esc} = \sqrt{6}\sigma_*$) is about a factor ${\sim}2$ smaller than ours (as computed in Eq.~\eqref{eqn:vesc}). Since the phase-space density goes as $\bar{F}_\mathrm{M} \propto v_{\mathrm{esc}}^{-3}$, this accounts for the difference. If instead we calculate the maximal coarse-grained phase-space density at the innermost radius, this difference reduces to a factor that is ${\sim}1$. This indicates that their assumptions regarding the dark matter density profile inside dwarf galaxies are somewhat at odds with the results of a higher-order Jeans analysis.

Some of the other references, e.g.~\cite{Gorbunov:2008ka}, follow a similar approach to~\cite{Boyarsky:2008ju}, while others estimate the dark matter velocity dispersion either by doing N-body simulations~\citep{Horiuchi:2013noa}, including a scaling parameter $\eta_* = \sigma_\mathrm{dm}/\sigma_\mathrm{stars}$~\citep{Dalcanton:2000hn} or calculating an \emph{average} quantity from a given profile/distribution function~\citep{Wang:2017hof, DiPaolo:2017geq}.

\subsubsection{Obtaining bounds}

In obtaining the phase-space bounds with Liouville's theorem, the authors of~\cite{Boyarsky:2008ju} estimated the phase-space density for Leo IV using maximal coarse-graining. In this work, we used both maximal and Gaussian coarse-graining, the latter leading to stronger bounds. At this point, it is worth noting the following: in this work, Gaussian coarse-graining results in a factor ${\sim}10$ larger phase-space density than the one used in~\cite{Boyarsky:2008ju} to obtain their reported bounds. On the other hand, this increase is partially compensated for by the factor ${\sim}4$ decrease in the maximum of the distribution function for resonantly produced sterile neutrinos (discussed in Sec.~\ref{sec:finegrainedcomparison}). The net result of this is that their bounds (using maximal coarse-graining) and our bounds (using Gaussian coarse-graining) are roughly similar.

Another important point of comparison is the way the bounds for resonantly produced sterile neutrinos are obtained. In~\cite{Boyarsky:2008ju} a vertical line in the $(m_{\mathrm{s}}, \sin^2 2\theta)$-plane is drawn at $m_\mathrm{s} = 1\,\mathrm{keV}$, independent of the mixing angle. For intermediate and small values of $\sin^2 2\theta$, this is a conservative bound, as the maximum of the distribution function decreases with smaller mixing angles. In Fig.~\ref{fig:RP_sterile}, we see that properly accounting for this effect, causes the bounds to extend to higher masses at those mixing angles. For larger mixing angles, however, we find that the maximum of the distribution function approaches that of a Fermi-Dirac distribution. As such, we would expect that the bounds weaken in this range of mixing angles. The absence of this feature is likely a result of the fact that this reference used distribution functions that do not go to low enough momenta. In addition, as we discuss extensively in Sec.~\ref{subsubsec:rp_sterile}, the numerical modelling of the distribution function in this region of parameter space is very challenging. With the current codes that are available, we have suggested that the only conservative option in this regime is to take the bound to be $m_{\mathrm{s}} \geq m_{\mathrm{FD}}$.

\vspace{-0.3cm}
\section{Conclusions}
\label{sec:Conclusions}

In this work we have improved and refined the lower bounds on the mass of fermionic dark matter from phase-space considerations. This was mainly achieved via a more detailed Jeans analysis that took advantage of higher-order moments of the velocity distribution, giving a significantly more constrained determination of the dark matter density profile in dwarf spheroidal galaxies. The very large phase-space density at the centre of these galaxies then allows us to put competitive constraints on both generic fermions, as well as specific fermionic dark matter candidates. Importantly, these are complementary to bounds from other probes, such as those from X-ray observations, which contain a very different set of astrophysical uncertainties.

In Sec.~\ref{sec:JeansAnalysis} we gave a review of the Jeans analysis, together with a summary of our approach to find the dark matter density and velocity dispersion at small radii. Sec.~\ref{sec:HowToObtainBounds} then laid out the method used to set phase-space bounds on fermionic dark matter from Pauli's principle and by using Liouville's theorem. In Sec.~\ref{sec:benchmark_models} we applied this formalism to a set of benchmark models, ranging from generic relativistically decoupled thermal fermions to sterile neutrinos produced through various production mechanisms. Finally, in Sec.~\ref{sec:Results} we presented our results and a comparison with previous studies. The constraints obtained in this work are summarised in Eq.~\eqref{eq:m_Pauli_result}, Table~\ref{tab:Liouville_bounds} and Figs.~\ref{fig:fmax_vs_mass},~\ref{fig:RP_sterile} and~\ref{fig:bounds_full_set}. Our main conclusions are:

\begin{itemize}
    \item Using the Pauli exclusion principle, our phase-space analysis yields a bound of: $m_{\mathrm{deg}} \geq 0.27^{+0.13}_{-0.09}\,\mathrm{keV}$ at 68\% CL and $m_{\mathrm{deg}} \geq 0.27^{+0.30}_{-0.14}\,\mathrm{keV}$ at 95\% CL.
    This constraint is the most robust, since it is independent of the fermionic particle physics model considered and the evolution of its distribution function. It also does not rely on the modelling of baryonic physics in the dwarf galaxies.
    \item Under the assumption that baryonic feedback does not increase the maximum of the distribution function, Liouville's theorem can further improve these bounds. We considered two approaches: maximal and Gaussian coarse-graining. For a range of benchmark models, including non-resonantly produced sterile neutrinos, we set a lower bound on the mass of the fermionic particle of $\mathcal{O}(1)$ keV (see Table~\ref{tab:Liouville_bounds}). 
    \item An extensive analysis regarding the maximum of the distribution function for resonantly produced sterile neutrinos has allowed us to place the most robust phase-space bounds on these particles. We also updated the complementary bound from Big Bang Nucleosynthesis. The results are summarised in Fig.~\ref{fig:RP_sterile}, where we compare them to other relevant astrophysical bounds.
\end{itemize}

In summary, the phase-space approach described in this work results in robust lower bounds on the mass of fermionic dark matter particles, with well-controlled errors and ranges of validity. This makes this method a valuable complement to other astrophysical probes.

We also provide an outlook as to how the bounds may be further improved in the future. Firstly, a better understanding of the evolution from a primordial distribution to the late-time state (e.g., the possible decrease of its maximum due to dynamical effects during virialisation, see e.g.~\cite{Shao:2012cg}) would allow one to make stronger statements regarding the inequality in Eq.~\eqref{eq:phase_space_bound_condition} than from Liouville's theorem. Secondly, for a given model, without a robust modelling of the primordial distribution function, one has to resort to the most conservative estimation of its maximum. As such, a more rigorous derivation of the primordial phase-space density in such scenarios will naturally strengthen the bounds presented here. Finally, we proposed a novel way of estimating the observed phase-space density by accessing information about the velocity dispersion of the dark matter. As we have mentioned previously, it is expected that the dark matter within dwarf galaxies with a low virial mass can potentially have relatively small velocity dispersions and large central densities. As such, a further improved modelling of the velocity-space distribution function can add more weight to the assumptions made in this work.

\vspace{-0.4cm}

\section*{Acknowledgements}
We are very grateful to Jacopo Ghiglieri and Francis-Yan Cyr-Racine for insightful discussions. We thank Andrew Robertson and Mark Lovell for producing dark matter beta profiles from the \texttt{Aquarius} suite and Josh Simon for providing the data for CVn I. JA is a recipient of an STFC quota studentship. NS is a recipient of a King's College London NMS Faculty Studentship. KB and AB are supported by the European Research Council (ERC) Advanced Grant ``NuBSM'' (694896). ME and MF are supported by the European Research Council under the European Union's Horizon 2020 program (ERC Grant Agreement No 648680 DARKHORIZONS). In addition, the work of MF was supported partly by the STFC Grant ST/P000258/1. ME is also supported by the Alexander von Humboldt Foundation. MO is supported by the STFC grant ST/R505134/1. The EDGE simulations were performed using the DiRAC Data Intensive service at Leicester, operated by the University of Leicester IT Services, which forms part of the STFC DiRAC High Performance Computer (HPC) Facility. The authors also made use of the Xmaris cluster at the Lorentz Institute and the Eureka HPC facility at the University of Surrey.

\section*{Data Availability Statement}
The data and code underlying this article are available on \texttt{GitHub} in the repository \texttt{FermionDSph} at \href{https://github.com/james-alvey-42/FermionDSph}{this URL}.
\vspace{-0.2cm}

\bibliographystyle{mnras}
\bibliography{fermdm} 

\begin{thebibliography}{}
\makeatletter
\relax
\def\mn@urlcharsother{\let\do\@makeother \do\$\do\&\do\#\do\^\do\_\do\%\do\~}
\def\mn@doi{\begingroup\mn@urlcharsother \@ifnextchar [ {\mn@doi@}
  {\mn@doi@[]}}
\def\mn@doi@[#1]#2{\def\@tempa{#1}\ifx\@tempa\@empty \href
  {http://dx.doi.org/#2} {doi:#2}\else \href {http://dx.doi.org/#2} {#1}\fi
  \endgroup}
\def\mn@eprint#1#2{\mn@eprint@#1:#2::\@nil}
\def\mn@eprint@arXiv#1{\href {http://arxiv.org/abs/#1} {{\tt arXiv:#1}}}
\def\mn@eprint@dblp#1{\href {http://dblp.uni-trier.de/rec/bibtex/#1.xml}
  {dblp:#1}}
\def\mn@eprint@#1:#2:#3:#4\@nil{\def\@tempa {#1}\def\@tempb {#2}\def\@tempc
  {#3}\ifx \@tempc \@empty \let \@tempc \@tempb \let \@tempb \@tempa \fi \ifx
  \@tempb \@empty \def\@tempb {arXiv}\fi \@ifundefined
  {mn@eprint@\@tempb}{\@tempb:\@tempc}{\expandafter \expandafter \csname
  mn@eprint@\@tempb\endcsname \expandafter{\@tempc}}}

\bibitem[\protect\citeauthoryear{Abada, Arcadi  \& Lucente}{Abada
  et~al.}{2014}]{Abada:2014zra}
Abada A.,  Arcadi G.,   Lucente M.,  2014, \mn@doi [JCAP]
  {10.1088/1475-7516/2014/10/001}, 10, 001

\bibitem[\protect\citeauthoryear{Abazajian}{Abazajian}{2006}]{Abazajian:2005gj}
Abazajian K.,  2006, \mn@doi [Phys. Rev. D] {10.1103/PhysRevD.73.063506}, 73,
  063506

\bibitem[\protect\citeauthoryear{Abazajian, Fuller  \& Tucker}{Abazajian
  et~al.}{2001}]{Abazajian:2001vt}
Abazajian K.,  Fuller G.~M.,   Tucker W.~H.,  2001, \mn@doi [Astrophys. J.]
  {10.1086/323867}, 562, 593

\bibitem[\protect\citeauthoryear{{Agertz} et~al.,}{{Agertz}
  et~al.}{2020}]{2020MNRAS.491.1656A}
{Agertz} O.,  et~al., 2020, \mn@doi [\mnras] {10.1093/mnras/stz3053}, \href
  {https://ui.adsabs.harvard.edu/abs/2020MNRAS.491.1656A} {491, 1656}

\bibitem[\protect\citeauthoryear{Aghanim et~al.}{Aghanim
  et~al.}{2018}]{Aghanim:2018eyx}
Aghanim N.,  et~al., 2018

\bibitem[\protect\citeauthoryear{Alvarez, Calore, Genina, Read, Serpico  \&
  Zaldivar}{Alvarez et~al.}{2020}]{Alvarez:2020cmw}
Alvarez A.,  Calore F.,  Genina A.,  Read J.,  Serpico P.~D.,   Zaldivar B.,
  2020, \mn@doi [JCAP] {10.1088/1475-7516/2020/09/004}, 09, 004

\bibitem[\protect\citeauthoryear{Amorisco \& Evans}{Amorisco \&
  Evans}{2012}]{Amorisco:2011hb}
Amorisco N.,  Evans N.,  2012, \mn@doi [Mon. Not. Roy. Astron. Soc.]
  {10.1111/j.1365-2966.2011.19684.x}, 419, 184

\bibitem[\protect\citeauthoryear{Anderhalden, Schneider, Maccio, Diemand  \&
  Bertone}{Anderhalden et~al.}{2013}]{Anderhalden:2012jc}
Anderhalden D.,  Schneider A.,  Maccio A.~V.,  Diemand J.,   Bertone G.,  2013,
  \mn@doi [JCAP] {10.1088/1475-7516/2013/03/014}, 03, 014

\bibitem[\protect\citeauthoryear{Asaka \& Shaposhnikov}{Asaka \&
  Shaposhnikov}{2005}]{Asaka:2005pn}
Asaka T.,  Shaposhnikov M.,  2005, \mn@doi [Phys. Lett. B]
  {10.1016/j.physletb.2005.06.020}, 620, 17

\bibitem[\protect\citeauthoryear{Asaka, Shaposhnikov  \& Kusenko}{Asaka
  et~al.}{2006}]{Asaka:2006ek}
Asaka T.,  Shaposhnikov M.,   Kusenko A.,  2006, \mn@doi [Phys. Lett. B]
  {10.1016/j.physletb.2006.05.067}, 638, 401

\bibitem[\protect\citeauthoryear{Aver, Olive  \& Skillman}{Aver
  et~al.}{2015}]{Aver:2015iza}
Aver E.,  Olive K.~A.,   Skillman E.~D.,  2015, \mn@doi [JCAP]
  {10.1088/1475-7516/2015/07/011}, 1507, 011

\bibitem[\protect\citeauthoryear{Bacon, Amara  \& Read}{Bacon
  et~al.}{2010}]{Bacon:2009aj}
Bacon D.,  Amara A.,   Read J.,  2010, \mn@doi [Mon. Not. Roy. Astron. Soc.]
  {10.1111/j.1365-2966.2010.17316.x}, 409, 389

\bibitem[\protect\citeauthoryear{Banik, Bovy, Bertone, Erkal  \& de Boer}{Banik
  et~al.}{2019}]{Banik:2019smi}
Banik N.,  Bovy J.,  Bertone G.,  Erkal D.,   de Boer T.,  2019

\bibitem[\protect\citeauthoryear{Baur, Palanque-Delabrouille, Yeche, Boyarsky,
  Ruchayskiy, Armengaud  \& Lesgourgues}{Baur et~al.}{2017}]{Baur:2017stq}
Baur J.,  Palanque-Delabrouille N.,  Yeche C.,  Boyarsky A.,  Ruchayskiy O.,
  Armengaud E.,   Lesgourgues J.,  2017, \mn@doi [JCAP]
  {10.1088/1475-7516/2017/12/013}, 12, 013

\bibitem[\protect\citeauthoryear{Bertone, Hooper  \& Silk}{Bertone
  et~al.}{2005}]{Bertone:2004pz}
Bertone G.,  Hooper D.,   Silk J.,  2005, \mn@doi [Phys. Rept.]
  {10.1016/j.physrep.2004.08.031}, 405, 279

\bibitem[\protect\citeauthoryear{{Binney} \& {Mamon}}{{Binney} \&
  {Mamon}}{1982}]{1982MNRAS.200..361B}
{Binney} J.,  {Mamon} G.~A.,  1982, \mn@doi [\mnras] {10.1093/mnras/200.2.361},
  \href {https://ui.adsabs.harvard.edu/abs/1982MNRAS.200..361B} {200, 361}

\bibitem[\protect\citeauthoryear{{Binney} \& {Tremaine}}{{Binney} \&
  {Tremaine}}{2008}]{2008gady.book.....B}
{Binney} J.,  {Tremaine} S.,  2008, {Galactic Dynamics: Second Edition}

\bibitem[\protect\citeauthoryear{Boehm et~al.}{Boehm
  et~al.}{2017}]{Boehm:2017wie}
Boehm C.,  et~al., 2017

\bibitem[\protect\citeauthoryear{Boyarsky, Ruchayskiy  \& Iakubovskyi}{Boyarsky
  et~al.}{2009a}]{Boyarsky:2008ju}
Boyarsky A.,  Ruchayskiy O.,   Iakubovskyi D.,  2009a, \mn@doi [JCAP]
  {10.1088/1475-7516/2009/03/005}, 03, 005

\bibitem[\protect\citeauthoryear{Boyarsky, Ruchayskiy  \&
  Shaposhnikov}{Boyarsky et~al.}{2009b}]{Boyarsky:2009ix}
Boyarsky A.,  Ruchayskiy O.,   Shaposhnikov M.,  2009b, \mn@doi [Ann. Rev.
  Nucl. Part. Sci.] {10.1146/annurev.nucl.010909.083654}, 59, 191

\bibitem[\protect\citeauthoryear{Boyarsky, Iakubovskyi  \& Ruchayskiy}{Boyarsky
  et~al.}{2012}]{Boyarsky:2012rt}
Boyarsky A.,  Iakubovskyi D.,   Ruchayskiy O.,  2012, \mn@doi [Phys. Dark
  Univ.] {10.1016/j.dark.2012.11.001}, 1, 136

\bibitem[\protect\citeauthoryear{Boyarsky, Ruchayskiy, Iakubovskyi  \&
  Franse}{Boyarsky et~al.}{2014}]{Boyarsky:2014jta}
Boyarsky A.,  Ruchayskiy O.,  Iakubovskyi D.,   Franse J.,  2014, \mn@doi
  [Phys. Rev. Lett.] {10.1103/PhysRevLett.113.251301}, 113, 251301

\bibitem[\protect\citeauthoryear{Boyarsky, Drewes, Lasserre, Mertens  \&
  Ruchayskiy}{Boyarsky et~al.}{2019}]{Boyarsky:2018tvu}
Boyarsky A.,  Drewes M.,  Lasserre T.,  Mertens S.,   Ruchayskiy O.,  2019,
  \mn@doi [Prog. Part. Nucl. Phys.] {10.1016/j.ppnp.2018.07.004}, 104, 1

\bibitem[\protect\citeauthoryear{Bulbul, Markevitch, Foster, Smith, Loewenstein
   \& Randall}{Bulbul et~al.}{2014}]{Bulbul:2014sua}
Bulbul E.,  Markevitch M.,  Foster A.,  Smith R.~K.,  Loewenstein M.,   Randall
  S.~W.,  2014, \mn@doi [Astrophys. J.] {10.1088/0004-637X/789/1/13}, 789, 13

\bibitem[\protect\citeauthoryear{Carr, Kuhnel  \& Sandstad}{Carr
  et~al.}{2016}]{Carr:2016drx}
Carr B.,  Kuhnel F.,   Sandstad M.,  2016, \mn@doi [Phys. Rev. D]
  {10.1103/PhysRevD.94.083504}, 94, 083504

\bibitem[\protect\citeauthoryear{Cherry \& Horiuchi}{Cherry \&
  Horiuchi}{2017}]{Cherry:2017dwu}
Cherry J.~F.,  Horiuchi S.,  2017, \mn@doi [Phys. Rev. D]
  {10.1103/PhysRevD.95.083015}, 95, 083015

\bibitem[\protect\citeauthoryear{Clowe, Bradac, Gonzalez, Markevitch, Randall,
  Jones  \& Zaritsky}{Clowe et~al.}{2006}]{Clowe:2006eq}
Clowe D.,  Bradac M.,  Gonzalez A.~H.,  Markevitch M.,  Randall S.~W.,  Jones
  C.,   Zaritsky D.,  2006, \mn@doi [Astrophys. J. Lett.] {10.1086/508162},
  648, L109

\bibitem[\protect\citeauthoryear{Dalcanton \& Hogan}{Dalcanton \&
  Hogan}{2001}]{Dalcanton:2000hn}
Dalcanton J.~J.,  Hogan C.~J.,  2001, \mn@doi [Astrophys. J.] {10.1086/323207},
  561, 35

\bibitem[\protect\citeauthoryear{Davoudiasl, Denton  \& McGady}{Davoudiasl
  et~al.}{2020}]{Davoudiasl:2020uig}
Davoudiasl H.,  Denton P.~B.,   McGady D.~A.,  2020

\bibitem[\protect\citeauthoryear{De~Gouvêa, Sen, Tangarife  \&
  Zhang}{De~Gouvêa et~al.}{2020}]{deGouvea:2019phk}
De~Gouvêa A.,  Sen M.,  Tangarife W.,   Zhang Y.,  2020, \mn@doi [Phys. Rev.
  Lett.] {10.1103/PhysRevLett.124.081802}, 124, 081802

\bibitem[\protect\citeauthoryear{Di~Paolo, Nesti  \& Villante}{Di~Paolo
  et~al.}{2018}]{DiPaolo:2017geq}
Di~Paolo C.,  Nesti F.,   Villante F.~L.,  2018, \mn@doi [Mon.\ Not.\ Roy.\
  Astron.\ Soc.] {10.1093/mnras/sty091}, 475, 5385

\bibitem[\protect\citeauthoryear{Dodelson}{Dodelson}{2003}]{Dodelson:2003ft}
Dodelson S.,  2003, {Modern Cosmology}.
Academic Press, Amsterdam

\bibitem[\protect\citeauthoryear{Dodelson}{Dodelson}{2011}]{Dodelson:2011qv}
Dodelson S.,  2011, \mn@doi [Int. J. Mod. Phys. D] {10.1142/S0218271811020561},
  20, 2749

\bibitem[\protect\citeauthoryear{Dodelson \& Widrow}{Dodelson \&
  Widrow}{1994}]{Dodelson:1993je}
Dodelson S.,  Widrow L.~M.,  1994, \mn@doi [Phys. Rev. Lett.]
  {10.1103/PhysRevLett.72.17}, 72, 17

\bibitem[\protect\citeauthoryear{Domcke \& Urbano}{Domcke \&
  Urbano}{2015}]{Domcke:2014kla}
Domcke V.,  Urbano A.,  2015, \mn@doi [JCAP] {10.1088/1475-7516/2015/01/002},
  01, 002

\bibitem[\protect\citeauthoryear{Drewes et~al.}{Drewes
  et~al.}{2017}]{Adhikari:2016bei}
Drewes M.,  et~al., 2017, \mn@doi [JCAP] {10.1088/1475-7516/2017/01/025}, 01,
  025

\bibitem[\protect\citeauthoryear{Efstathiou, Schaye  \& Theuns}{Efstathiou
  et~al.}{2000}]{Efstathiou:2000kk}
Efstathiou G.,  Schaye J.,   Theuns T.,  2000, \mn@doi [Phil. Trans. Roy. Soc.
  Lond. A] {10.1098/rsta.2000.0629}, 358, 2049

\bibitem[\protect\citeauthoryear{Fern\'andez, Terlevich, Díaz, Terlevich  \&
  Rosales-Ortega}{Fern\'andez et~al.}{2018}]{Fern_ndez_2018}
Fern\'andez V.,  Terlevich E.,  Díaz A.~I.,  Terlevich R.,   Rosales-Ortega
  F.~F.,  2018, \mn@doi [Monthly Notices of the Royal Astronomical Society]
  {10.1093/mnras/sty1206}, 478, 5301–5319

\bibitem[\protect\citeauthoryear{Flewelling et~al.,}{Flewelling
  et~al.}{2019}]{flewelling2019panstarrs1}
Flewelling H.~A.,  et~al., 2019, The Pan-STARRS1 Database and Data Products
  (\mn@eprint {arXiv} {1612.05243})

\bibitem[\protect\citeauthoryear{Frigerio \& Yaguna}{Frigerio \&
  Yaguna}{2015}]{Frigerio:2014ifa}
Frigerio M.,  Yaguna C.~E.,  2015, \mn@doi [Eur. Phys. J. C]
  {10.1140/epjc/s10052-014-3252-1}, 75, 31

\bibitem[\protect\citeauthoryear{Garzilli, Boyarsky  \& Ruchayskiy}{Garzilli
  et~al.}{2017}]{Garzilli:2015iwa}
Garzilli A.,  Boyarsky A.,   Ruchayskiy O.,  2017, \mn@doi [Phys. Lett. B]
  {10.1016/j.physletb.2017.08.022}, 773, 258

\bibitem[\protect\citeauthoryear{Garzilli, Ruchayskiy, Magalich  \&
  Boyarsky}{Garzilli et~al.}{2019a}]{Garzilli:2019qki}
Garzilli A.,  Ruchayskiy O.,  Magalich A.,   Boyarsky A.,  2019a

\bibitem[\protect\citeauthoryear{Garzilli, Magalich, Theuns, Frenk, Weniger,
  Ruchayskiy  \& Boyarsky}{Garzilli et~al.}{2019b}]{Garzilli:2018jqh}
Garzilli A.,  Magalich A.,  Theuns T.,  Frenk C.~S.,  Weniger C.,  Ruchayskiy
  O.,   Boyarsky A.,  2019b, \mn@doi [Mon. Not. Roy. Astron. Soc.]
  {10.1093/mnras/stz2188}, 489, 3456

\bibitem[\protect\citeauthoryear{{Genina} et~al.,}{{Genina}
  et~al.}{2018}]{2018MNRAS.474.1398G}
{Genina} A.,  et~al., 2018, \mn@doi [\mnras] {10.1093/mnras/stx2855}, \href
  {https://ui.adsabs.harvard.edu/abs/2018MNRAS.474.1398G} {474, 1398}

\bibitem[\protect\citeauthoryear{Gerhard}{Gerhard}{1993}]{Gerhard:1993tn}
Gerhard O.,  1993, Mon. Not. Roy. Astron. Soc., 265, 213

\bibitem[\protect\citeauthoryear{Ghiglieri \& Laine}{Ghiglieri \&
  Laine}{2015}]{Ghiglieri:2015jua}
Ghiglieri J.,  Laine M.,  2015, \mn@doi [JHEP] {10.1007/JHEP11(2015)171}, 11,
  171

\bibitem[\protect\citeauthoryear{Ghiglieri \& Laine}{Ghiglieri \&
  Laine}{2019}]{Ghiglieri:2019kbw}
Ghiglieri J.,  Laine M.,  2019, \mn@doi [JHEP] {10.1007/JHEP07(2019)078}, 07,
  078

\bibitem[\protect\citeauthoryear{Ghiglieri \& Laine}{Ghiglieri \&
  Laine}{2020}]{Ghiglieri:2020ulj}
Ghiglieri J.,  Laine M.,  2020

\bibitem[\protect\citeauthoryear{Gilman, Birrer, Nierenberg, Treu, Du  \&
  Benson}{Gilman et~al.}{2020}]{Gilman:2019nap}
Gilman D.,  Birrer S.,  Nierenberg A.,  Treu T.,  Du X.,   Benson A.,  2020,
  \mn@doi [Mon. Not. Roy. Astron. Soc.] {10.1093/mnras/stz3480}, 491, 6077

\bibitem[\protect\citeauthoryear{Gorbunov, Khmelnitsky  \& Rubakov}{Gorbunov
  et~al.}{2008}]{Gorbunov:2008ka}
Gorbunov D.,  Khmelnitsky A.,   Rubakov V.,  2008, \mn@doi [JCAP]
  {10.1088/1475-7516/2008/10/041}, 10, 041

\bibitem[\protect\citeauthoryear{Helmi}{Helmi}{2020}]{Helmi:2020otr}
Helmi A.,  2020, \mn@doi [Ann. Rev. Astron. Astrophys.]
  {10.1146/annurev-astro-032620-021917}, 58, annurev

\bibitem[\protect\citeauthoryear{Horiuchi, Humphrey, Onorbe, Abazajian,
  Kaplinghat  \& Garrison-Kimmel}{Horiuchi et~al.}{2014}]{Horiuchi:2013noa}
Horiuchi S.,  Humphrey P.~J.,  Onorbe J.,  Abazajian K.~N.,  Kaplinghat M.,
  Garrison-Kimmel S.,  2014, \mn@doi [Phys. Rev. D]
  {10.1103/PhysRevD.89.025017}, 89, 025017

\bibitem[\protect\citeauthoryear{Hui, Ostriker, Tremaine  \& Witten}{Hui
  et~al.}{2017}]{Hui:2016ltb}
Hui L.,  Ostriker J.~P.,  Tremaine S.,   Witten E.,  2017, \mn@doi [Phys. Rev.
  D] {10.1103/PhysRevD.95.043541}, 95, 043541

\bibitem[\protect\citeauthoryear{Ibata, Lewis  \& Irwin}{Ibata
  et~al.}{2002}]{Ibata:2001iv}
Ibata R.,  Lewis G.,   Irwin M.,  2002, \mn@doi [Mon. Not. Roy. Astron. Soc.]
  {10.1046/j.1365-8711.2002.05358.x}, 332, 915

\bibitem[\protect\citeauthoryear{Izotov, Thuan  \& Guseva}{Izotov
  et~al.}{2014}]{Izotov:2014fga}
Izotov Y.,  Thuan T.,   Guseva N.,  2014, \mn@doi [Mon. Not. Roy. Astron. Soc.]
  {10.1093/mnras/stu1771}, 445, 778

\bibitem[\protect\citeauthoryear{{Jeans}}{{Jeans}}{1922}]{1922MNRAS..82..122J}
{Jeans} J.~H.,  1922, \mn@doi [\mnras] {10.1093/mnras/82.3.122}, \href
  {https://ui.adsabs.harvard.edu/abs/1922MNRAS..82..122J} {82, 122}

\bibitem[\protect\citeauthoryear{Jethwa, Erkal  \& Belokurov}{Jethwa
  et~al.}{2018}]{Jethwa:2016gra}
Jethwa P.,  Erkal D.,   Belokurov V.,  2018, \mn@doi [Mon. Not. Roy. Astron.
  Soc.] {10.1093/mnras/stx2330}, 473, 2060

\bibitem[\protect\citeauthoryear{Johnston, Spergel  \& Haydn}{Johnston
  et~al.}{2002}]{Johnston:2001wh}
Johnston K.~V.,  Spergel D.~N.,   Haydn C.,  2002, \mn@doi [Astrophys. J.]
  {10.1086/339791}, 570, 656

\bibitem[\protect\citeauthoryear{Kim, Peter  \& Hargis}{Kim
  et~al.}{2018}]{Kim:2017iwr}
Kim S.~Y.,  Peter A. H.~G.,   Hargis J.~R.,  2018, \mn@doi [Phys. Rev. Lett.]
  {10.1103/PhysRevLett.121.211302}, 121, 211302

\bibitem[\protect\citeauthoryear{Koposov, Irwin, Belokurov, Gonzales-Solares,
  Yoldas, Lewis, Metcalfe  \& Schanks}{Koposov et~al.}{2014}]{Koposov:2014lja}
Koposov S.,  Irwin M.,  Belokurov V.,  Gonzales-Solares E.,  Yoldas A.~K.,
  Lewis J.,  Metcalfe N.,   Schanks T.,  2014, \mn@doi [Mon. Not. Roy. Astron.
  Soc.] {10.1093/mnrasl/slu060}, 442, 85

\bibitem[\protect\citeauthoryear{Kusenko}{Kusenko}{2006}]{Kusenko:2006rh}
Kusenko A.,  2006, \mn@doi [Phys. Rev. Lett.] {10.1103/PhysRevLett.97.241301},
  97, 241301

\bibitem[\protect\citeauthoryear{König, Merle  \& Totzauer}{König
  et~al.}{2016}]{Konig:2016dzg}
König J.,  Merle A.,   Totzauer M.,  2016, \mn@doi [JCAP]
  {10.1088/1475-7516/2016/11/038}, 11, 038

\bibitem[\protect\citeauthoryear{Laine \& Shaposhnikov}{Laine \&
  Shaposhnikov}{2008}]{Laine:2008pg}
Laine M.,  Shaposhnikov M.,  2008, \mn@doi [JCAP]
  {10.1088/1475-7516/2008/06/031}, 06, 031

\bibitem[\protect\citeauthoryear{Lokas \& Mamon}{Lokas \&
  Mamon}{2003}]{Lokas:2003ks}
Lokas E.~L.,  Mamon G.~A.,  2003, \mn@doi [Mon. Not. Roy. Astron. Soc.]
  {10.1046/j.1365-8711.2003.06684.x}, 343, 401

\bibitem[\protect\citeauthoryear{Lokas, Mamon  \& Prada}{Lokas
  et~al.}{2005}]{Lokas:2004sw}
Lokas E.~L.,  Mamon G.~A.,   Prada F.,  2005, \mn@doi [Mon. Not. Roy. Astron.
  Soc.] {10.1111/j.1365-2966.2005.09497.x}, 363, 918

\bibitem[\protect\citeauthoryear{Lovell, Frenk, Eke, Jenkins, Gao  \&
  Theuns}{Lovell et~al.}{2014}]{Lovell:2013ola}
Lovell M.~R.,  Frenk C.~S.,  Eke V.~R.,  Jenkins A.,  Gao L.,   Theuns T.,
  2014, \mn@doi [Mon. Not. Roy. Astron. Soc.] {10.1093/mnras/stt2431}, 439, 300

\bibitem[\protect\citeauthoryear{Lovell et~al.,}{Lovell
  et~al.}{2016}]{Lovell:2015psz}
Lovell M.~R.,  et~al., 2016, \mn@doi [Mon. Not. Roy. Astron. Soc.]
  {10.1093/mnras/stw1317}, 461, 60

\bibitem[\protect\citeauthoryear{Marsh}{Marsh}{2016}]{Marsh:2015xka}
Marsh D. J.~E.,  2016, \mn@doi [Phys. Rept.] {10.1016/j.physrep.2016.06.005},
  643, 1

\bibitem[\protect\citeauthoryear{Mateo, Olszewski  \& Walker}{Mateo
  et~al.}{2008}]{Mateo:2007xh}
Mateo M.,  Olszewski E.~W.,   Walker M.~G.,  2008, \mn@doi [Astrophys. J.]
  {10.1086/522326}, 675, 201

\bibitem[\protect\citeauthoryear{McConnachie}{McConnachie}{2012}]{McConnachie:2012vd}
McConnachie A.~W.,  2012, \mn@doi [Astron. J.] {10.1088/0004-6256/144/1/4},
  144, 4

\bibitem[\protect\citeauthoryear{McMonigal et~al.,}{McMonigal
  et~al.}{2014}]{McMonigal:2014naa}
McMonigal B.,  et~al., 2014, \mn@doi [Mon. Not. Roy. Astron. Soc.]
  {10.1093/mnras/stu1659}, 444, 3139

\bibitem[\protect\citeauthoryear{Menci, Sanchez, Castellano  \& Grazian}{Menci
  et~al.}{2016}]{Menci:2016eww}
Menci N.,  Sanchez N.,  Castellano M.,   Grazian A.,  2016, \mn@doi [Astrophys.
  J.] {10.3847/0004-637X/818/1/90}, 818, 90

\bibitem[\protect\citeauthoryear{Menci, Merle, Totzauer, Schneider, Grazian,
  Castellano  \& Sanchez}{Menci et~al.}{2017}]{Menci:2017nsr}
Menci N.,  Merle A.,  Totzauer M.,  Schneider A.,  Grazian A.,  Castellano M.,
   Sanchez N.~G.,  2017, \mn@doi [Astrophys. J.] {10.3847/1538-4357/836/1/61},
  836, 61

\bibitem[\protect\citeauthoryear{{Merrifield} \& {Kent}}{{Merrifield} \&
  {Kent}}{1990}]{1990AJ.....99.1548M}
{Merrifield} M.~R.,  {Kent} S.~M.,  1990, \mn@doi [\aj] {10.1086/115438}, \href
  {https://ui.adsabs.harvard.edu/abs/1990AJ.....99.1548M} {99, 1548}

\bibitem[\protect\citeauthoryear{Mertens et~al.,}{Mertens
  et~al.}{2015}]{Mertens:2014nha}
Mertens S.,  et~al., 2015, \mn@doi [JCAP] {10.1088/1475-7516/2015/02/020}, 02,
  020

\bibitem[\protect\citeauthoryear{Nadler et~al.}{Nadler
  et~al.}{2020}]{Nadler:2020prv}
Nadler E.,  et~al., 2020

\bibitem[\protect\citeauthoryear{Nemevsek, Senjanovic  \& Zhang}{Nemevsek
  et~al.}{2012}]{Nemevsek:2012cd}
Nemevsek M.,  Senjanovic G.,   Zhang Y.,  2012, \mn@doi [JCAP]
  {10.1088/1475-7516/2012/07/006}, 07, 006

\bibitem[\protect\citeauthoryear{Palanque-Delabrouille, Yèche, Schöneberg,
  Lesgourgues, Walther, Chabanier  \& Armengaud}{Palanque-Delabrouille
  et~al.}{2020}]{Palanque-Delabrouille:2019iyz}
Palanque-Delabrouille N.,  Yèche C.,  Schöneberg N.,  Lesgourgues J.,
  Walther M.,  Chabanier S.,   Armengaud E.,  2020, \mn@doi [JCAP]
  {10.1088/1475-7516/2020/04/038}, 04, 038

\bibitem[\protect\citeauthoryear{Pardo \& Spergel}{Pardo \&
  Spergel}{2020}]{Pardo:2020epc}
Pardo K.,  Spergel D.~N.,  2020

\bibitem[\protect\citeauthoryear{Peebles}{Peebles}{1984}]{Peebles:1984zz}
Peebles P.,  1984, \mn@doi [Astrophys. J.] {10.1086/161714}, 277, 470

\bibitem[\protect\citeauthoryear{Peimbert, Peimbert  \& Luridiana}{Peimbert
  et~al.}{2016}]{Peimbert:2016bdg}
Peimbert A.,  Peimbert M.,   Luridiana V.,  2016, Rev. Mex. Astron. Astrofis.,
  52, 419

\bibitem[\protect\citeauthoryear{Peirani, Durier  \&
  De~Freitas~Pacheco}{Peirani et~al.}{2006}]{Peirani:2005kw}
Peirani S.,  Durier F.,   De~Freitas~Pacheco J.~A.,  2006, \mn@doi [Mon. Not.
  Roy. Astron. Soc.] {10.1111/j.1365-2966.2006.10149.x}, 367, 1011

\bibitem[\protect\citeauthoryear{Piattella, Rodrigues, Fabris  \& de
  Freitas~Pacheco}{Piattella et~al.}{2013}]{Piattella:2013cma}
Piattella O.~F.,  Rodrigues D.~C.,  Fabris J.~C.,   de Freitas~Pacheco J.~A.,
  2013, \mn@doi [JCAP] {10.1088/1475-7516/2013/11/002}, 11, 002

\bibitem[\protect\citeauthoryear{Pitrou, Coc, Uzan  \& Vangioni}{Pitrou
  et~al.}{2018}]{Pitrou:2018cgg}
Pitrou C.,  Coc A.,  Uzan J.-P.,   Vangioni E.,  2018, \mn@doi [Phys. Rept.]
  {10.1016/j.physrep.2018.04.005}, 754, 1

\bibitem[\protect\citeauthoryear{Read, Iorio, Agertz  \& Fraternali}{Read
  et~al.}{2017}]{Read:2017lvq}
Read J.,  Iorio G.,  Agertz O.,   Fraternali F.,  2017, \mn@doi [Mon. Not. Roy.
  Astron. Soc.] {10.1093/mnras/stx147}, 467, 2019

\bibitem[\protect\citeauthoryear{{Read}, {Walker}  \& {Steger}}{{Read}
  et~al.}{2018}]{2018MNRAS.481..860R}
{Read} J.~I.,  {Walker} M.~G.,   {Steger} P.,  2018, \mn@doi [\mnras]
  {10.1093/mnras/sty2286}, \href
  {https://ui.adsabs.harvard.edu/abs/2018MNRAS.481..860R} {481, 860}

\bibitem[\protect\citeauthoryear{Read, Walker  \& Steger}{Read
  et~al.}{2019a}]{Read:2018fxs}
Read J.,  Walker M.,   Steger P.,  2019a, \mn@doi [Mon.\ Not.\ Roy.\ Astron.\
  Soc.] {10.1093/mnras/sty3404}, 484, 1401

\bibitem[\protect\citeauthoryear{{Read}, {Walker}  \& {Steger}}{{Read}
  et~al.}{2019b}]{2019MNRAS.484.1401R}
{Read} J.~I.,  {Walker} M.~G.,   {Steger} P.,  2019b, \mn@doi [\mnras]
  {10.1093/mnras/sty3404}, \href
  {https://ui.adsabs.harvard.edu/abs/2019MNRAS.484.1401R} {484, 1401}

\bibitem[\protect\citeauthoryear{Richardson \& Fairbairn}{Richardson \&
  Fairbairn}{2013}]{Richardson:2012ig}
Richardson T.,  Fairbairn M.,  2013, \mn@doi [Mon. Not. Roy. Astron. Soc.]
  {10.1093/mnras/stt686}, 432, 3361

\bibitem[\protect\citeauthoryear{Richardson \& Fairbairn}{Richardson \&
  Fairbairn}{2014}]{Richardson:2014mra}
Richardson T.,  Fairbairn M.,  2014, \mn@doi [Mon. Not. Roy. Astron. Soc.]
  {10.1093/mnras/stu691}, 441, 1584

\bibitem[\protect\citeauthoryear{Rubin \& Ford}{Rubin \&
  Ford}{1970}]{Rubin:1970zza}
Rubin V.~C.,  Ford W.Kent J.,  1970, \mn@doi [Astrophys. J.] {10.1086/150317},
  159, 379

\bibitem[\protect\citeauthoryear{Savchenko \& Rudakovskyi}{Savchenko \&
  Rudakovskyi}{2019}]{Savchenko:2019qnn}
Savchenko D.,  Rudakovskyi A.,  2019, \mn@doi [Mon. Not. Roy. Astron. Soc.]
  {10.1093/mnras/stz1573}, 487, 5711

\bibitem[\protect\citeauthoryear{Shao, Gao, Theuns  \& Frenk}{Shao
  et~al.}{2013}]{Shao:2012cg}
Shao S.,  Gao L.,  Theuns T.,   Frenk C.~S.,  2013, \mn@doi [Mon. Not. Roy.
  Astron. Soc.] {10.1093/mnras/stt053}, 430, 2346

\bibitem[\protect\citeauthoryear{Shaposhnikov \& Tkachev}{Shaposhnikov \&
  Tkachev}{2006}]{Shaposhnikov:2006xi}
Shaposhnikov M.,  Tkachev I.,  2006, \mn@doi [Phys. Lett. B]
  {10.1016/j.physletb.2006.06.063}, 639, 414

\bibitem[\protect\citeauthoryear{Shi \& Fuller}{Shi \&
  Fuller}{1999}]{Shi:1998km}
Shi X.-D.,  Fuller G.~M.,  1999, \mn@doi [Phys. Rev. Lett.]
  {10.1103/PhysRevLett.82.2832}, 82, 2832

\bibitem[\protect\citeauthoryear{{Simon}}{{Simon}}{2019}]{Simon2019xd}
{Simon} J.~D.,  2019, \mn@doi [\araa] {10.1146/annurev-astro-091918-104453},
  \href {https://ui.adsabs.harvard.edu/abs/2019ARA&A..57..375S} {57, 375}

\bibitem[\protect\citeauthoryear{Simon \& Geha}{Simon \&
  Geha}{2007}]{Simon:2007dq}
Simon J.~D.,  Geha M.,  2007, \mn@doi [Astrophys. J.] {10.1086/521816}, 670,
  313

\bibitem[\protect\citeauthoryear{Spencer, Mateo, Walker  \& Olszewski}{Spencer
  et~al.}{2017}]{Spencer_2017}
Spencer M.~E.,  Mateo M.,  Walker M.~G.,   Olszewski E.~W.,  2017, \mn@doi [The
  Astrophysical Journal] {10.3847/1538-4357/836/2/202}, 836, 202

\bibitem[\protect\citeauthoryear{Spencer, Mateo, Olszewski, Walker, McConnachie
   \& Kirby}{Spencer et~al.}{2018}]{Spencer_2018}
Spencer M.~E.,  Mateo M.,  Olszewski E.~W.,  Walker M.~G.,  McConnachie A.~W.,
   Kirby E.~N.,  2018, \mn@doi [The Astronomical Journal]
  {10.3847/1538-3881/aae3e4}, 156, 257

\bibitem[\protect\citeauthoryear{Springel et~al.,}{Springel
  et~al.}{2008}]{Springel:2008cc}
Springel V.,  et~al., 2008, \mn@doi [Mon. Not. Roy. Astron. Soc.]
  {10.1111/j.1365-2966.2008.14066.x}, 391, 1685

\bibitem[\protect\citeauthoryear{{Teyssier}}{{Teyssier}}{2002}]{2002A&A...385..337T}
{Teyssier} R.,  2002, \mn@doi [\aap] {10.1051/0004-6361:20011817}, \href
  {http://adsabs.harvard.edu/abs/2002A%26A...385..337T} {385, 337}

\bibitem[\protect\citeauthoryear{Tremaine \& Gunn}{Tremaine \&
  Gunn}{1979}]{Tremaine:1979we}
Tremaine S.,  Gunn J.,  1979, \mn@doi [Phys. Rev. Lett.]
  {10.1103/PhysRevLett.42.407}, 42, 407

\bibitem[\protect\citeauthoryear{Valerdi, Peimbert, Peimbert  \&
  Sixtos}{Valerdi et~al.}{2019}]{Valerdi:2019beb}
Valerdi M.,  Peimbert A.,  Peimbert M.,   Sixtos A.,  2019, \mn@doi [Astrophys.
  J.] {10.3847/1538-4357/ab14e4}, 876, 98

\bibitem[\protect\citeauthoryear{Vegetti, Despali, Lovell  \& Enzi}{Vegetti
  et~al.}{2018}]{Vegetti:2018dly}
Vegetti S.,  Despali G.,  Lovell M.,   Enzi W.,  2018, \mn@doi [Mon. Not. Roy.
  Astron. Soc.] {10.1093/mnras/sty2393}, 481, 3661

\bibitem[\protect\citeauthoryear{Venumadhav, Cyr-Racine, Abazajian  \&
  Hirata}{Venumadhav et~al.}{2016}]{Venumadhav:2015pla}
Venumadhav T.,  Cyr-Racine F.-Y.,  Abazajian K.~N.,   Hirata C.~M.,  2016,
  \mn@doi [Phys. Rev. D] {10.1103/PhysRevD.94.043515}, 94, 043515

\bibitem[\protect\citeauthoryear{Viel, Becker, Bolton  \& Haehnelt}{Viel
  et~al.}{2013}]{Viel:2013fqw}
Viel M.,  Becker G.~D.,  Bolton J.~S.,   Haehnelt M.~G.,  2013, \mn@doi [Phys.
  Rev. D] {10.1103/PhysRevD.88.043502}, 88, 043502

\bibitem[\protect\citeauthoryear{Vogelsberger et~al.,}{Vogelsberger
  et~al.}{2009}]{Vogelsberger:2008qb}
Vogelsberger M.,  et~al., 2009, \mn@doi [Mon. Not. Roy. Astron. Soc.]
  {10.1111/j.1365-2966.2009.14630.x}, 395, 797

\bibitem[\protect\citeauthoryear{Walker \& Penarrubia}{Walker \&
  Penarrubia}{2011}]{Walker:2011zu}
Walker M.~G.,  Penarrubia J.,  2011, \mn@doi [Astrophys. J.]
  {10.1088/0004-637X/742/1/20}, 742, 20

\bibitem[\protect\citeauthoryear{Walker, Mateo  \& Olszewski}{Walker
  et~al.}{2009}]{Walker:2008ax}
Walker M.~G.,  Mateo M.,   Olszewski E.,  2009, \mn@doi [Astron. J.]
  {10.1088/0004-6256/137/2/3100}, 137, 3100

\bibitem[\protect\citeauthoryear{Walker, Olszewski  \& Mateo}{Walker
  et~al.}{2015}]{Walker_2015}
Walker M.~G.,  Olszewski E.~W.,   Mateo M.,  2015, \mn@doi [Monthly Notices of
  the Royal Astronomical Society] {10.1093/mnras/stv099}, 448, 2717–2732

\bibitem[\protect\citeauthoryear{Wang, Cherry, Horiuchi  \& Strigari}{Wang
  et~al.}{2017}]{Wang:2017hof}
Wang M.-Y.,  Cherry J.~F.,  Horiuchi S.,   Strigari L.~E.,  2017

\bibitem[\protect\citeauthoryear{White}{White}{1994}]{White:1994bn}
White S.~D.,  1994, in {Les Houches Summer School on Cosmology and Large Scale
  Structure (Session 60)}. pp 349--430 (\mn@eprint {arXiv} {astro-ph/9410043})

\bibitem[\protect\citeauthoryear{Wolf, Martinez, Bullock, Kaplinghat, Geha,
  Munoz, Simon  \& Avedo}{Wolf et~al.}{2010}]{Wolf:2009tu}
Wolf J.,  Martinez G.~D.,  Bullock J.~S.,  Kaplinghat M.,  Geha M.,  Munoz
  R.~R.,  Simon J.~D.,   Avedo F.~F.,  2010, \mn@doi [Mon. Not. Roy. Astron.
  Soc.] {10.1111/j.1365-2966.2010.16753.x}, 406, 1220

\bibitem[\protect\citeauthoryear{Zurek}{Zurek}{2014}]{Zurek:2013wia}
Zurek K.~M.,  2014, \mn@doi [Phys. Rept.] {10.1016/j.physrep.2013.12.001}, 537,
  91

\makeatother
\end{thebibliography}

\bsp
\label{lastpage}

\end{document}